\documentclass[12pt,english,floatfix,showkeys,superscriptaddress,aps,prd]{revtex4}
\usepackage[utf8]{inputenc}
\usepackage[english]{babel}
\usepackage[T1]{fontenc}
\usepackage{lmodern}
\usepackage{amsmath}
\usepackage{subfigure}
\usepackage{amssymb}
\usepackage{graphicx}
\usepackage{float}
\usepackage{esint}
\usepackage{longtable}
\usepackage{dcolumn}
\usepackage{babel}
\usepackage{csquotes}
\usepackage{color}
\usepackage{slashed}
\usepackage{simplewick}
\usepackage[utf8]{inputenc}                   
\usepackage{amsmath,latexsym}

\usepackage{hyperref}
\hypersetup{
    colorlinks,
    citecolor=blue,
    filecolor=green,
    linkcolor=purple,
    urlcolor=red,
}

\usepackage{slashed}

\usepackage{hyperref}
\hypersetup{colorlinks,breaklinks,
			citecolor=[rgb]{0,0.0,1.0},
            urlcolor=[rgb]{0.0,0.0,1.0},
            linkcolor=[rgb]{0,0.5,0.9}}

\begin{document}

\title{Implications of Complexity Factor on Evolution of New Dynamical and Static Wormholes in $f(R, T)$ Gravity}

\author{M. Zubair}
\email{mzubairkk@gmail.com; drmzubair@cuilahore.edu.pk}
\affiliation{Department of Mathematics, COMSATS University Islamabad, Lahore-Campus, Lahore, Pakistan}
\affiliation{National Astronomical Observatories, Chinese Academy of Sciences, Beijing 100101, China}
\author{Hina Azmat}
\email{hinaazmat0959@gmail.com}
\affiliation{Department of Mathematics, COMSATS University Islamabad, Lahore-Campus, Lahore, Pakistan}
\author{Quratulien Muneer}
\email{anie.muneer@gmail.com}
\affiliation{Department of Mathematics, COMSATS University Islamabad, Lahore-Campus, Lahore, Pakistan}
\author{Saira Waheed}
\email{swaheed@pmu.edu.sa}
\affiliation{Prince Mohammad Bin Fahd University, Al Khobar, 31952 Kingdom of Saudi Arabia}
\author{M. B. Cruz}
\email{messiasdebritocruz@servidor.uepb.edu.br}
\affiliation{Universidade Estadual da Para\'iba (UEPB), \\ Centro de Ci\^encias Exatas e Sociais Aplicadas (CCEA), \\ R. Alfredo Lustosa Cabral, s/n, Salgadinho, Patos - PB, 58706-550 - Brazil.}

\begin{abstract}
This study presents new spherically symmetric and dynamical wormhole solutions supported by ordinary matter modeled as an anisotropic fluid, exhibiting a traversable nature. To achieve this goal, we adopt different approaches to obtain both evolving static and genuinely dynamical solutions, such as imposing a viable condition on the Ricci scalar, considering an anisotropic equation of state, and choosing a suitable energy density profile. For each derived shape function, we analyze the corresponding $2D$ and $3D$ embedding diagrams and verify their compatibility with the weak energy condition through density plots. The equilibrium conditions are also explored graphically to assess the stability of the obtained solutions, which are shown to be stable within the analyzed framework. Additionally, we investigate the complexity factor associated with each configuration, examining its dependence on both temporal evolution and the coupling parameter $\lambda$ of the $f(R,T)$ theory.
\end{abstract}

 
\maketitle

\date{\today}

\section{Introduction}
General Relativity (GR) is widely regarded as the most accurate and well-established theory for describing gravitational phenomena \cite{Einstein:1916vd}. Among its most intriguing predictions are black holes (BHs) \cite{Oppenheimer:1939ue} and wormholes (WHs) \cite{Hawking:1988ae, Morris:1988cz}. The existence of BHs has been confirmed through direct observations—most notably through the detection of gravitational waves (GWs) resulting from the merger of binary BH systems, with remarkable precision achieved by the LIGO and Virgo collaborations \cite{LIGOScientific:2016aoc}. Furthermore, the images of black hole shadows captured by the Event Horizon Telescope (EHT) collaboration have provided striking visual confirmation of the existence and properties of these compact objects, offering strong support for GR's predictions \cite{EventHorizonTelescope:2019dse}. In contrast, while wormholes remain undetected, they continue to be a fascinating theoretical consequence of GR, attracting significant interest due to their potential as space-time tunnels connecting distant regions of the universe \cite{Einstein:1935tc, Morris:1988tu, Frolov:2023res}.

Wormholes are hypothetical tunnel-like structures in spacetime, capable of connecting distant regions of the universe—or even points belonging to entirely different universes. The idea was first introduced through the geometric interpretation of the Schwarzschild solution by Flamm \cite{Flamm:1916}, and later revisited by Einstein and Rosen \cite{Einstein:1935tc}, who proposed a geometric construction connecting two black holes, now known as the Einstein-Rosen bridge. In recent years, the construction of wormhole geometries in various gravitational scenarios has garnered increasing attention. Such solutions are generally classified into two main categories: Euclidean wormholes and Lorentzian wormholes \cite{Coleman:1988cy, Giddings:1988wv}. In a seminal study, Morris and Thorne proposed the existence of Lorentzian wormholes that are stable, traversable, static, and spherically symmetric within the framework of GR \cite{Morris:1988cz}. Moreover, they explored the possibility of traversal through these hypothetical tunnels by analyzing their fundamental properties and the physical conditions required for their viability.

In this context, the crucial property of wormhole traversability imposes significant constraints, such as the requirement of exotic matter or the absence of event horizons \cite{Morris:1988cz, Visser:2003yf, Hochberg:1998ha, MontelongoGarcia:2010xd}. In particular, traversable WHs supported by exotic matter exhibit negative energy density and pressure, which leads to the violation of the Null Energy Condition (NEC). One of the main challenges in constructing wormhole geometries, therefore, lies in ensuring their compatibility with the established energy conditions. Despite the uncertainties surrounding their physical viability, WHs continue to attract considerable interest, especially due to their potential role at the interface between GR and Quantum Mechanics (QM) \cite{Li:2008sw, Sengupta:2023yof, Muniz:2024jzg, Cruz:2024ihb, deSSilva:2024gdc}. In regimes of extremely intense gravitational fields, it is expected that new properties of spacetime may emerge, requiring a more comprehensive description than that provided by GR alone. In this scenario, non-trivial structures may arise at microscopic scales, as suggested by theoretical models developed within the frameworks of Loop Quantum Gravity (LQG) \cite{Rovelli:2014ssa, Rovelli:2003wd} and String Theory (ST) \cite{Zwiebach:2004tj}.

In recent decades, several studies have explored WHs solutions within the context of modified gravity theories, aiming to avoid the use of exotic matter and, thus, preserve the NEC. Promising results have shown that non-minimal coupling between matter and geometry can mitigate—or even eliminate—NEC violation, especially at the wormhole throat. Recent investigations have examined WHs in various gravitational scenarios, including: $f(R)$ gravity \cite{Tanaka:2007xm, Lobo:2009ip, DeBenedictis:2012qz}, $f(Q)$ gravity \cite{Parsaei:2022wnu, Hassan:2022jgn, Rastgoo:2024udl}, $f(R,T)$ gravity \cite{Sokoliuk:2022, Sokoliuk:2022epjc, Errehymy:2023ann, Magalhaes:2023prd, Sadeghi:2022mpl, Gogoi:2023jcap}, teleparallel gravity \cite{Singh:2020rai, Parsaei:2022wnu, Sokoliuk:2022efj, Li:2022vtn}, Gauss-Bonnet gravity \cite{ZeeshanGul:2024vem, Sharif:2020xxj}, Brans-Dicke theory \cite{Papantonopoulos:2019ugr, Bhadra:2005is, Tretyakova:2015vaa}, and higher-dimensional models \cite{Torii:2013xba, Baruah:2019cfg}. The $f(R,T)$ gravity has stood out due to the adoption of different shape functions and equations of state, including models with phantom energy and charged configurations \cite{Harko:2011kv, Zubair:2016cde}. Variations such as $f(R^2,T)$ have also been studied, resulting in numerically obtained solutions \cite{Sahoo:2017ual}. Conformal Killing symmetry and the Karmarkar condition have been employed to derive physically viable solutions \cite{Errehymy:2024spg}. Furthermore, there is growing interest in dynamic wormholes, which can be obtained without exotic matter by embedding the wormhole metric into a Friedmann–Lemaître–Robertson–Walker (FLRW) spacetime \cite{Pan:2014lua, Pan:2014oaa}, thereby connecting these solutions to the cosmological background.

Moreover, relevant studies have been conducted with the aim of understanding thermal corrections in dynamic wormhole solutions \cite{Rehman:2020myc}. For instance, Ref. \cite{Mehdizadeh:2021kgv} investigated dynamic wormholes in the context of particle creation within non-equilibrium thermodynamics, with particular emphasis on the role of temperature. In cosmological scenarios based on power-law and exponential models, aspects such as temperature and surface gravity associated with the shape functions were also analyzed \cite{Bahamonde:2016ixz, Cataldo:2008ku, Lobo:2012qq, Kuhfittig:2013hva}.

This work aims to deepen the investigation of wormhole geometries by proposing new static and dynamic solutions within the framework of $f(R,T)$ modified gravity. The central objective is to examine how the inclusion of the Ricci scalar $R$ and the trace of the energy-momentum tensor 
$T$ in the gravitational action influences the formation and characteristics of such structures. The underlying hypothesis is that the non-minimal coupling between matter and geometry, inherent to $f(R,T)$ gravity, may enable the existence of wormhole geometries without the need for exotic matter—or at least with a significant attenuation of energy condition violations. Special attention is devoted to a careful analysis of the null, weak, and dominant energy conditions, with the goal of identifying the regimes in which these conditions can be satisfied or relaxed. By addressing both static and dynamic configurations, this study seeks to contribute to a better understanding of the role of modified gravity theories in supporting traversable wormholes and their broader implications for alternative models of gravitation.

This article is organized as follows: In Section \ref{f_gravity}, we briefly present the theoretical framework of modified gravity $f(R, T)$. In Section \ref{f_fields_equations}, we obtain dynamic wormhole solutions by imposing a specific condition on the Ricci scalar. Section \ref{WHs_ES} is dedicated to the analysis of static solutions based on an anisotropic equation of state, while in Section \ref{WH_EDP} we investigate static solutions motivated by a physically inspired energy density profile. In Section \ref{equi_cond}, we analyze the hydrostatic equilibrium condition using the Tolman–Oppenheimer–Volkoff (TOV) equation applied to the obtained solutions. Then, in Section \ref{y_factor}, we discuss the complexity factor associated with the proposed wormhole geometries. Finally, in Section \ref{conclusions}, we present our conclusions and perspectives for future research. Throughout this paper, we will utilize natural units with $G=c=1$ and adopt the metric signature $(-,+,+,+)$.

\section{The $f(R, T)$ Gravity} \label{f_gravity}
In this section, we present brief introduction to theoretical model of modified gravity $f(R,T)$, which will be considered in our analysis. In recent years, the $f(R,T)$ theory has emerged as a promising approach within the framework of gravitation, proving successful in the investigation of various cosmological aspects \cite{Harko:2011kv}. Its action is given by:
\begin{eqnarray}\label{f_action}
    S &=& \frac{1}{16 \pi} \int f(R,T)\sqrt{g} d^4x + \int L_{m} \sqrt{-g} d^4x,
\end{eqnarray}
where $f(R,T)$ is an arbitrary function of the Ricci scalar $R$ and the trace of the energy-momentum tensor $T=g^{\mu \nu}T_{\mu \nu}$. The term $L_m$ denotes the matter Lagrangian density.

By applying the variation of the gravitational field action $S$, Eq. \eqref{f_action}, with respect to the metric tensor $g_{\mu \nu}$, we obtain the following relation:
\begin{eqnarray}\label{field_equations}
    8\pi T_{\mu \nu} - f_{T}(R,T)T_{\mu \nu} - f_{T}(R,T) \Theta_{\mu \nu} &=& f_{R}(R,T)R_{\mu \nu}
    - \frac{1}{2}f(R,T)g_{\mu \nu} \nonumber \\ &+& (g_{\mu \nu} \Box - \nabla_{\mu} \nabla_{\nu}) f_{R}(R,T).
\end{eqnarray}
Here, $\nabla$ and $\Box$ denote the covariant derivative and the d'Alembert operator, respectively. The notations $f_R(R,T)$ and $f_T(R,T)$ refer to the partial derivatives of the generic function $f(R,T)$ with respect to $R$ and $T$, respectively. Additionally, the term $\Theta_{ij}$ appearing in Eq. \eqref{field_equations} is defined as:
\begin{eqnarray}\label{energy_tensor}
    \Theta_{\mu \nu} &=& g^{\alpha \beta} \frac{\delta T_{\alpha \beta}}{\delta g^{\mu \nu}} \\ &=& -2T_{\mu \nu} + g_{\mu \nu}L_{m} - 2 g^{\alpha \beta} \frac{\partial^{2}L_{m}}{\partial
    g^{\mu \nu} \partial g^{\alpha \beta}}.
\end{eqnarray}

For our purposes, we assume that the energy-momentum tensor takes the following form: 
\begin{eqnarray}\label{energy_moment_tensor}
    T_{\mu \nu} = \text{diag}(\rho, -p_r, -p_t, -p_t),
\end{eqnarray}
where $\rho$ denotes the energy density, while $p_r$ and $p_t$ correspond, respectively, to the radial and transverse components of pressure. Moreover, we consider $L_m=-P$, where $P$ denotes the total pressure, expressed as
\begin{eqnarray} \label{pressure}
    P = \frac{(p_r+2p_t)}{3} .
\end{eqnarray}
Thus, by substituting Eq. \eqref{pressure} into Eq. \eqref{energy_moment_tensor}, the variation of the energy-momentum tensor takes the following form:
\begin{eqnarray}
    \Theta_{\mu \nu} = -2T_{\mu \nu} - P g_{\mu \nu} .
\end{eqnarray}

In general, as can be seen from Eq. \eqref{field_equations}, the field equations also depend, through the tensor $\Theta_{\mu \nu}$, on the physical nature of the matter field. Thus, in $f(R,T)$ gravity, depending on the choice of the matter source, different functional forms of $f$ can lead to distinct theoretical models, each associated with a specific matter configuration. For the purposes of the present analysis, we consider a particular class of modified gravity defined by 
\begin{eqnarray} \label{chosen_f}
    f(R,T) = R + \lambda T,
\end{eqnarray}
where $\lambda$ denotes the coupling constant characterizing the interaction between matter and geometry. Therefore, by substituting Eq. \eqref{chosen_f} into Eq. \eqref{field_equations}, we obtain the following field equations:
\begin{equation} \label{final_field_equations}
    R_{\mu \nu} - \frac{1}{2}Rg_{\mu \nu} = (8 \pi + \lambda)T_{\mu \nu} + \frac{\lambda}{2}(\rho - P) g_{\mu \nu }.
\end{equation}

With the modified field equations established, we now proceed to their application in the analysis of dynamical wormhole metrics within the framework of $f(R,T)$ gravity.

\section{Dynamical Wormholes in $f(R,T)$ Gravity: Field Equations and Solutions} \label{f_fields_equations}
In this section, we aim to determine the dynamic wormhole solution by solving the field equations, Eq. \eqref{final_field_equations}. Our goal is to explicitly obtain the shape and redshift functions that characterize this geometry, as well as to highlight the assumptions adopted to make the analysis feasible.

We begin by considering the line element that describes the geometry of a spherically symmetric, traversable, and dynamic wormhole \cite{Cataldo:2008ku}, given by
\begin{eqnarray}\label{form_WH_metric}
ds^2 &=& e^{2\Phi(r)}dt^2-a(t)^{2}\bigg[\frac{dr^2}{1-b(r)/r}+r^{2}(d\theta^{2}+sin^{2}\theta d\Phi^{2})\bigg],
\end{eqnarray}
where $a(t)$ is the scale factor, $\Phi(r)$ represents the redshift function, and $b(r)$ denotes the shape function. These generic functions, which depend only on the radial coordinate $r$, define the main characteristics of the wormhole geometry. For the solution to be physically viable, the shape function must satisfy the conditions established in \cite{Morris:1988cz}. Moreover, it is well known that, for a wormhole geometry with zero tidal force, the redshift function must remain finite, which implies $\Phi(r) = 0$ \cite{Zubair:2022jjm}.

Therefore, since our goal is to investigate the existence of dynamical wormholes with anisotropic pressure using a non-static metric that asymptotically approaches the static FLRW metric, we adopt the condition $\Phi(r) = 0$ throughout this analysis. From Eqs. \eqref{final_field_equations} and \eqref{form_WH_metric}, we obtain the following system of field equations:
\begin{eqnarray} \label{set_field_equations}
    3H^2 + \frac{b'}{a^2 r^2} &=& (8 \pi + \lambda) \rho + \frac{\lambda}{2}\left(\rho - \frac{p_{r} + 2p_{t}}{3}\right), \\\label{set_field_equation_1}
    2 \dot{H} + 3H^2 + \frac{b}{a^2 r^3 } &=& (8 \pi + \lambda)(-p_{r}) + \frac{\lambda}{2}\left(\rho -\frac{p_{r}+2p_{t}}{3}\right), \\\label{set_field_equation_2}
    2 \dot{H} + 3H^2 + \frac{1}{2r^2 a^2}\left(b^{'} - \frac{b}{r}\right) &=& (8 \pi + \lambda)(-p_{t})
    + \frac{\lambda}{2}\left(\rho - \frac{p_{r}+2p_{t}}{3}\right).
\end{eqnarray}
Here, the prime symbol ($'$) denotes a derivative with respect to the radial coordinate $r$, while the dot ($\dot{\ }$) represents a derivative with respect to time $t$. Thus, by defining the Hubble parameter as $H = \dot{a}(t)/a(t)$, and using the system of equations given by Eqs. \eqref{set_field_equations}, \eqref{set_field_equation_1}, and \eqref{set_field_equation_2}, we derive the following expressions for the energy density $\rho$, the radial pressure $p_r$, and the tangential pressure $p_t$:
\begin{eqnarray} \label{set_ro_pre}
    \rho(r,t) &=& \frac{9 a^2 H^2(\lambda + 8 \pi)r^2 - 3a^2 \dot{H}\lambda r^2 + 4b^{'}(\lambda + 6 \pi)}{6 a^2(\lambda + 4 \pi)(\lambda + 8 \pi)r^2}, \\\label{set_ro_pre_1}
    p_r(r,t) &=& - \frac{9 a^2H^2 \lambda r^3 + 72 \pi a^2H^2r^3 + 3a^2 \dot{H}(3 \lambda + 16 \pi)r^3 + 6 b(\lambda + 4 \pi) - 2b^{'} \lambda r}{6 a^2(\lambda + 4 \pi (\lambda + 8 \pi)r^3}, \\\label{set_ro_pre_2}
    p_t(r,t) &=& -\frac{r(9 a^2 H^2(\lambda + 8 \pi)r^2 + b^{'}(\lambda + 12 \pi))
    + 3a^2\dot{H}(3 \lambda + 16 \pi)r^3 - 3 b(\lambda + 4 \pi)}{6 a^2(\lambda + 4 \pi)(\lambda + 8 \pi)r^3}.
\end{eqnarray}

As previously mentioned, for traversable wormholes in four dimensions, energy conditions must be violated in order to ensure the flaring-out condition near the throat \cite{Hochberg:1997wp}. However, later studies have shown that within the context of higher-dimensional theories, these conditions can be satisfied at the throat radius \cite{Mazharimousavi:2010bm, Mehdizadeh:2015dta, Dehghani:2009zza}. Therefore, for the existence of a dynamical wormhole solution, it is expected that the Weak Energy Condition (WEC) is satisfied. The WEC is defined by the following inequalities: $\rho \geq 0$, $p_r + \rho \geq 0$ and $p_t + \rho \geq 0$. It is worth noting that the last two inequalities correspond to the NEC. From the first inequality, together with Eqs. \eqref{set_ro_pre}, \eqref{set_ro_pre_1} and \eqref{set_ro_pre_2}, we obtain:
\begin{eqnarray} \label{set_ro+pressure}
    \rho + p_r = \frac{-2 a^2 \dot{H} r^3 - b + b^{'}r}{a^2 (\lambda +8 \pi) r^3}, \\\label{set_ro+pressure_1}
    \rho + p_t = \frac{-4 a^2 \dot{H} r^3 + b + b^{'} r}{2 a^2 (\lambda +8 \pi) r^3}.
\end{eqnarray}

\subsection*{Cosmological Wormhole}

\begin{figure}[h!]
\centering
\includegraphics[width=0.4\textwidth]{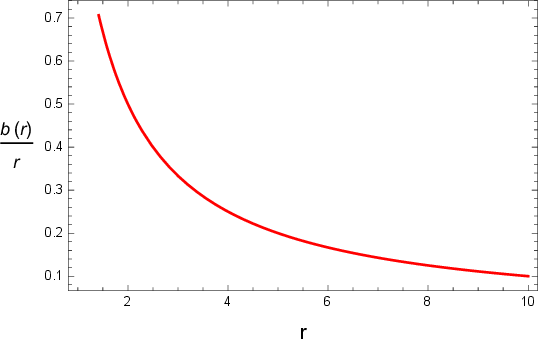}
\includegraphics[width=0.38\textwidth]{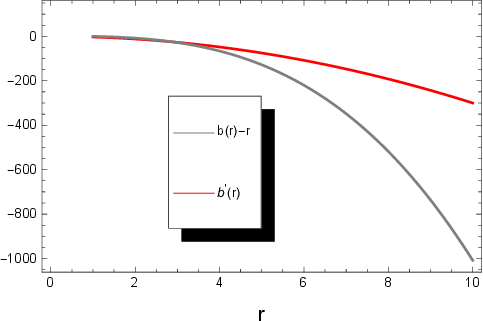}
\includegraphics[width=0.4\textwidth]{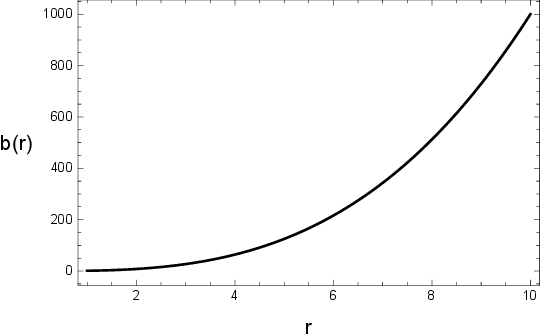}
\caption{Behavior of the shape function $b(r)$ for the cases $C_1 = 0$, $1$, and $-1$. The upper left plot illustrates the behavior of $b(r)/r$ for $C_1 = 0$, while the plot on the right shows the function $b(r)$ for $C_1 = -1$. Finally, the bottom plot presents the behavior of the shape function for $C_1 = 1$. In all cases, we have adopted $r_0 = 1$.}
\label{figb}
\end{figure}

\begin{figure}[h!]
\centering
\includegraphics[width=0.4\textwidth]{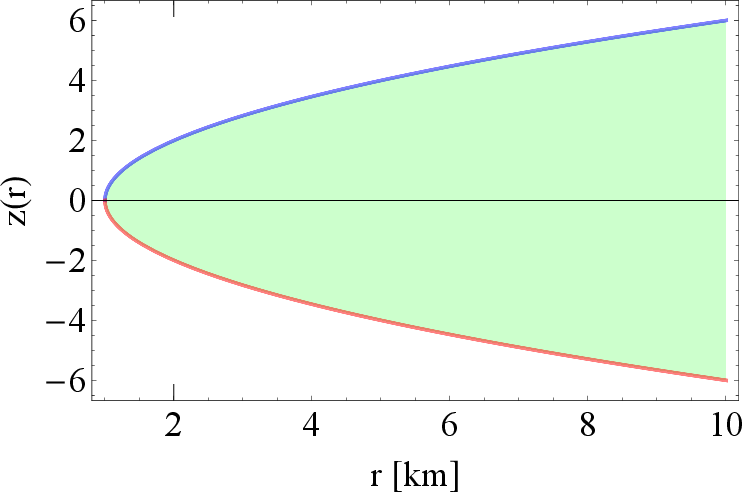}
\includegraphics[width=0.4\textwidth]{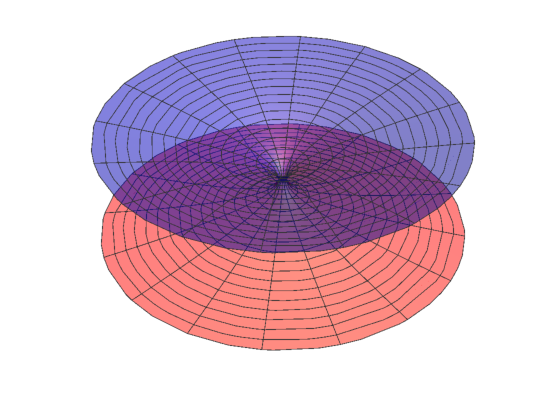}
\caption{The behavior of the two-dimensional embedding diagram is shown in the left plot, while the right plot provides a complete visualization of the dynamical wormhole for the case $C_1 = 0$.}
\label{figeb}
\end{figure}

In this subsection, we propose the construction of cosmological wormholes by assuming a specific condition on the Ricci scalar. From the set of equations \eqref{set_ro_pre}, \eqref{set_ro_pre_1}, and \eqref{set_ro_pre_2}, we observe that the system consists of three equations and five unknowns, namely: $\rho(r,t)$, $p_r(r,t)$, $p_t(r,t)$, $b(r)$ and $a(t)$. To investigate the geometry of these wormholes in a cosmological context — for which several approaches are available in the literature \cite{Sushkov:2005kj, Lobo:2005us, Anchordoqui:1996jh} — we adopt a condition on the Ricci scalar, $R$, which plays a key role in constructing the dynamical solution. In this case, the Ricci scalar takes the form:
\begin{eqnarray} \label{scale_R_r_t}
    R(t,r) = -6(\dot{H}+2H^2) - \frac{2b^{'}}{a^2r^2}.
\end{eqnarray}
Here, we observe that, due to the second term, the Ricci scalar explicitly depends on the radial coordinate $r$. However, as mentioned earlier, we impose a specific condition on the Ricci scalar that is often adopted in the construction of dynamical wormholes in cosmological contexts: we assume that $R$ is a function of time only, that is, $R = R(t)$. This choice is motivated by the fact that, in cosmological models based on the FLRW metric, the curvature scalar depends solely on cosmic time, reflecting the large-scale homogeneity and isotropy of the universe. By adopting this assumption — that is, that $R$ is independent of $r$ — we obtain the following expression for the shape function:
\begin{eqnarray} \label{b_function_1}
    b(r) = C_1 r^3 + C_2,
\end{eqnarray}
where $C_1$ and $C_2$ are integration constants. By applying the boundary condition $b(r_0) = r_0$ to Eq. \eqref{b_function_1}, we determine the value of $C_2$, and consequently, the shape function takes the following form:
\begin{eqnarray} \label{b_function_2}
    b(r) = C_1r^3 + r_0 \left(1 - C_1 r_0^2 \right).
\end{eqnarray}
Furthermore, by imposing the condition $b'(r_0) < 1$ in Eq. \eqref{b_function_2}, we obtain that $C_1 < 1/3r_0^2$. For $C_1 = 0$, corresponding to a flat universe, the shape function $b(r)$ satisfies the asymptotically flat condition, that is, $b(r)/r \rightarrow 0$ as $r \rightarrow \infty$. In the case of $C_1 = -1$, associated with an open universe, the condition $b'(r_0) < 1$ also remains satisfied. On the other hand, for $C_1 = 1$, which represents a closed universe, the function $b(r)$ becomes strictly increasing, as illustrated in Fig. \ref{figb}. The embedding diagram of the dynamical wormhole in the flat case ($C_1 = 0$) is shown in Fig. \ref{figeb}.

By substituting Eq. \eqref{b_function_2} into Eqs. \eqref{set_ro_pre}, \eqref{set_ro_pre_1}, and \eqref{set_ro_pre_2}, we obtain the following expressions:
\begin{eqnarray} \label{set_ro_cb}
    \rho(r,t) &=& \rho_{cb}(t) = \frac{a^2(3H^2(\lambda+8\pi) - \dot{H}\lambda) + 4C_1(\lambda+6\pi)}{2a^2(\lambda+4\pi)(\lambda+8\pi)}, \\\label{set_ro_cb_1}
    p_r(r,t) &=& p_{cb}(t) - \frac{C_1(8\pi(r^3-r_{0}^3) - 2\lambda r_{0}^3) + 2(\lambda+4\pi)r_0}{2a^2(\lambda+4\pi)(\lambda + 8\pi)r^3}, \\\label{set_ro_cb_12}
    p_t(r,t) &=& p_{cb}(t) - \frac{C_1(\lambda+12\pi)r^3 - (\lambda+4\pi)(C_1r^3-r_0(C_1r_{0}^{2}-1))}{2a^2(\lambda+4\pi)(\lambda+8\pi)r^3},
\end{eqnarray}
where $p_{cb}$ represents the pressure component associated with the cosmological background, given by: 
\begin{eqnarray}
    p_{cb} = -\frac{H^2(24\pi+3\lambda) +(16 \pi + 3 \lambda) \dot{H}}{2(\lambda + 4 \pi)(\lambda + 8 \pi)} .
\end{eqnarray}
In the following, we explore the properties of the evolving dynamical wormhole by examining the scale factor for three distinct cases: $C_1=0$, $C_1=1$ and $-1$. To determine it, we employ a linear equation of state (EOS), $p_{cb}=\omega \rho_{cb}$, in the cosmological background. Under this assumption, the following differential equation is obtained:
\begin{eqnarray}\label{cosmological_differential_equation}
    a^2 \left(3H^2(\lambda+8\pi)(\omega+1) + \dot{H}(\lambda(-\omega) + 3 \lambda + 16 \pi)\right) + 4 C_1 (\lambda+6\pi) \omega = 0.
\end{eqnarray}

\subsubsection*{Case I: $C_1 = 0$}

We begin with Eq. \eqref{cosmological_differential_equation} for the case $\omega = 1$, which corresponds to decelerated cosmic phases. Accordingly, we obtain the following expression for $a(t)$:
\begin{eqnarray} \label{11*}
    a(t) = (A+3t)^{1/3} B ,
\end{eqnarray}
where $A$ and $B$ are integration constants. To verify the validity of the WEC for the WH solution, we rewrite the expressions for $\rho(t)$, $\rho(t)+p_r(r,t)$, and $\rho(t)+p_t(r,t)$ as follows:
\begin{eqnarray} \label{set_eqs_C1_0}
    \rho(t) &=& \frac{3}{(\lambda + 8 \pi (A+3t)^2}, \\\label{set_eqs_C1_0_1}
    \rho(t) + p_r(r,t) &=& \frac{6 B^2 r^3-r_0(A+3 t)^{4/3}}{B^2(\lambda + 8 \pi)r^3(A+3t)^2}, \\\label{set_eqs_C1_0_2}
    \rho(t) + p_t(r,t) &=& \frac{r_0 (A+3 t)^{4/3}+12 B^2 r^3}{2B^2(\lambda + 8 \pi)r^3(A+3t)^2}.
\end{eqnarray}

\begin{figure}[h!]
\centering
\includegraphics[width=0.32\textwidth]{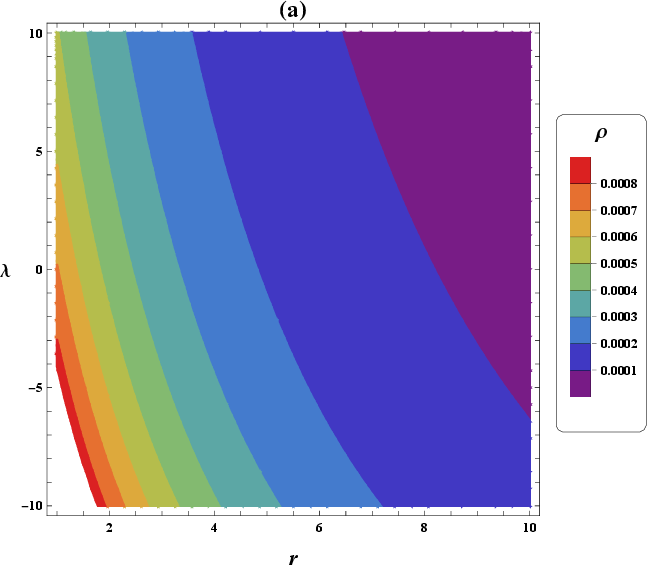}
\includegraphics[width=0.32\textwidth]{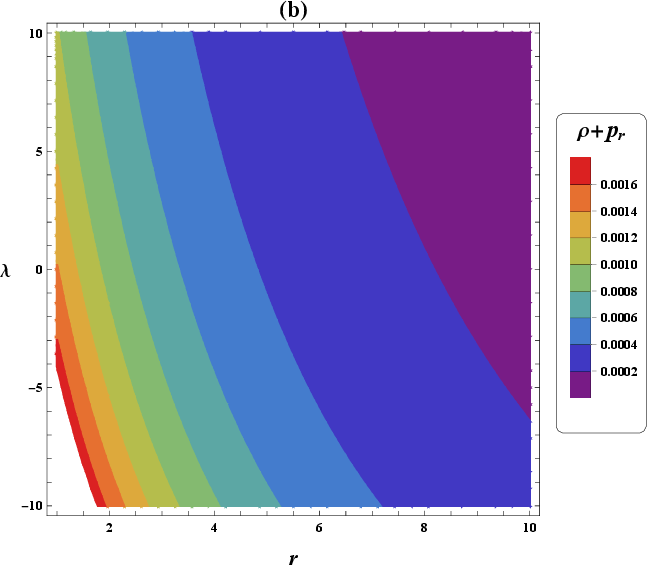}
\includegraphics[width=0.32\textwidth]{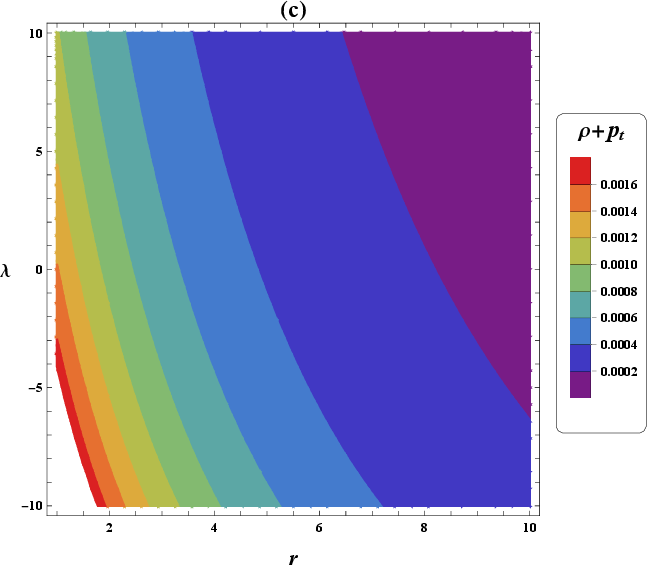}
\caption{Evolution of the WEC over time $t$, considering a specific range of $\lambda$ for the case where $C_1 = 0$. The adopted parameters are: $A = 10$, $B = 10$, $r = 5$, and $r_0 = 1$.}
\label{fig0}
\end{figure}

For a better understanding of the results, we use Eqs. \eqref{set_eqs_C1_0}, \eqref{set_eqs_C1_0_1}, and \eqref{set_eqs_C1_0_2} to plot the graphs shown in Fig. \ref{fig0}, which correspond to the energy constraints, enabling the visualization of the existence of dynamic WHs. Furthermore, to investigate their evolution, we analyze the validity of the WEC at the throat, considering both radial and temporal coordinates. This allows us to identify the regions of the parameter space that satisfy this condition. Graphical analysis reveals that: $\rho \geq 0$ is satisfied for $\{\lambda \geq -25,\ \forall A\}$, $\rho + p_r \geq 0$ is valid for $\{\lambda \geq -25,\ A \geq 0,\ \forall B\}$ and $\rho + p_t \geq 0$ also holds for $\{\lambda \geq -25,\ A \geq 0,\ \forall B\}$. Therefore, it can be concluded that dynamical wormholes can exist in a flat universe within the framework of $f(R,T)$ gravity.

\subsubsection*{Case II: $C_1=-1$}

At this stage, we revisit Eq. \eqref{cosmological_differential_equation} to investigate the behavior of the scale factor $a(t)$ in an open cosmological background, i.e., for the case $C_1=-1$. By considering 
$\omega = -1$, which characterizes an inflationary expansion regime, the equation reduces to the following differential equation:
\begin{eqnarray} \label{differential_eq_c1_-1}
    (\lambda + 4 \pi)a(t) \ddot{a}(t) + (\lambda+4 \pi)\dot{a}(t)^2+(\lambda+6 \pi) = 0.
\end{eqnarray}
The solution to the above equation for $a(t)$ can be expressed as:
\begin{eqnarray} \nonumber
    a(t) &=& \pm \frac{1}{\sqrt{\lambda^3+14 \pi \lambda^2+64 \pi^2 \lambda+96 \pi^3}} \Bigg{[}(e^{2a_1(\lambda + 4 \pi)} -144 \pi^3 t^2-84 \pi^2 \lambda t^2-16 \pi \lambda^2 t^2 \label{a_C_-1}\\ &-& \lambda^3 t^2-288 \pi^3 a_2 t -168 \pi^2 a_2\lambda t -32 \pi a_2 \lambda^2 t - a_2^2 \lambda^3-16 \pi a_2^2 \lambda^2 -84 \pi^2 a_2^2 \lambda \\ &+& 2 a_2 \lambda^3 t-144 \pi^3 a_2^2)\Bigg{]}^{1/2} \nonumber,
\end{eqnarray}
where $a_1$ and $a_2$ are integration constants. By substituting the expression of $a(t)$ with a positive sign and $b(r)$ from Eq. \eqref{b_function_2} into Eqs. \eqref{set_ro_cb}, \eqref{set_ro_cb_1}, and \eqref{set_ro_cb_12}, we obtain:
\begin{eqnarray} \label{set_eqs_C1_-1_1}
    \rho(t) = \frac{(\lambda +6 \pi)^2(8(\lambda +6 \pi)^2(\lambda^2+9 \pi \lambda+20 \pi^2)
    (a_2+t)^2-(3\lambda+16 \pi)e^{2a_1(\lambda+4 \pi)})}{2(\lambda+8 \pi)(e^{2a_1(\lambda+4 \pi)} -(\lambda +4 \pi)(\lambda+6 \pi)^2(a_2+t)^2)^2} ,
\end{eqnarray}
\begin{eqnarray} \label{set_eqs_C1_-1_2}
    \rho(t) + p_r(r,t) &=& \frac{(\lambda+6 \pi)}{2(\lambda+8 \pi)(e^{2 a_1(\lambda+4 \pi)}-(\lambda+4 \pi)(\lambda 6 \pi)^2 (a_2+t)^2)^2}\Bigg{\{}(\lambda+6 \pi)(8(\lambda + 6\pi)^2 \nonumber \\ &\times& (\lambda^2+9 \pi \lambda+20 \pi^2)(a_2+t)^2 -(3\lambda+16 \pi)e^{2 a_1 (\lambda+4 \pi)})-\frac{1}{r^3}e^{2 a_1(\lambda+4 \pi)}\Bigg{[}(\lambda^2 \nonumber \\ &\times& (2(r_0^3+r_0)-3 r^3)+2 \pi \lambda(8(r_0^3+r_0)-21 r^3)-32 \pi^2(4r^3 -r_0(r_0^2+1))) \nonumber \\ &+& 2(\lambda+4 \pi (\lambda+6 \pi)^2(a_2+t)^2(8 \pi^2(5r^3-2(r_0^3+r_0))+8 \pi \lambda(r^3-r_0(r_0^2 \nonumber \\ &+& 1)) + \lambda^2 (-r_0(r_0^2+1))\Bigg{]}\Bigg{\}} ,
\end{eqnarray}
\begin{eqnarray} \label{set_eqs_C1_-1_3}
    \rho(t)+p_t(r,t) &=& \frac{(\lambda+6 \pi)}{2(\lambda+8 \pi)(e^{2a_1(\lambda+4 \pi)}-(\lambda+4 \pi(\lambda +6 \pi)^2(a_2+t)^2)^2}\Bigg{\{}(\lambda+6 \pi)(8(\lambda \nonumber \\ &+& 6 \pi)^2(\lambda^2+9 \pi \lambda
    +20 \pi^2)(a_2+t)^2-(3 \lambda+16 \pi) e^{2a_1(\lambda+4 \pi)})-\frac{1}{r^3}\Bigg{[}(\lambda \nonumber \\ &+& 4 \pi)(\lambda+6 \pi)^2(a_2+t)^2 (8 \pi \lambda(2r^3+r_0^3+r_0)+16 \pi^2(5r^3+r_0^3+r_0) \nonumber \\ &+& \lambda^2 r_0(r_0^2+1))-e^{2a_1(\lambda+4 \pi)}(\lambda^2 (3 r^3+r_0^3+r_0)+2 \pi \lambda(21r^3+4(r_0^3+r_0)) \nonumber \\ &+& 16 \pi^2 (8r^3+r_0^3+r_0))\Bigg{]}\Bigg{\}} .
\end{eqnarray}

\begin{figure}[h!]
\centering
\includegraphics[width=0.32\textwidth]{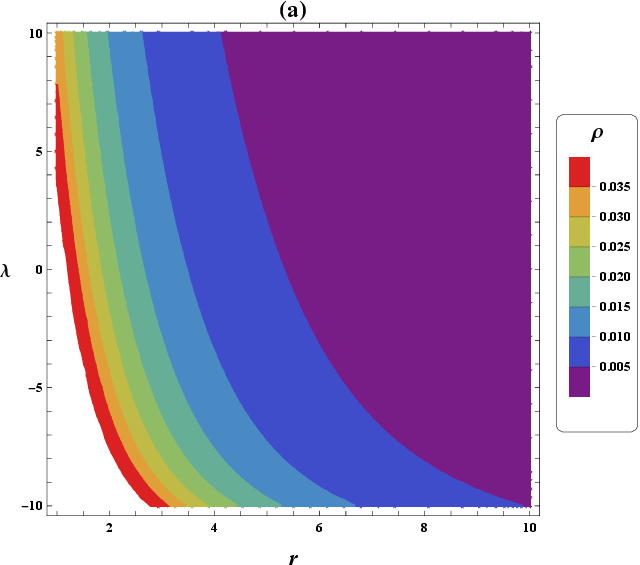}
\includegraphics[width=0.32\textwidth]{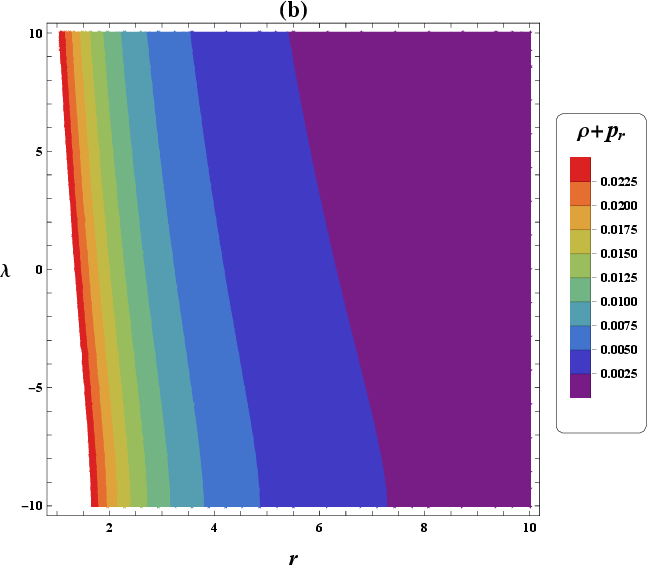}
\includegraphics[width=0.32\textwidth]{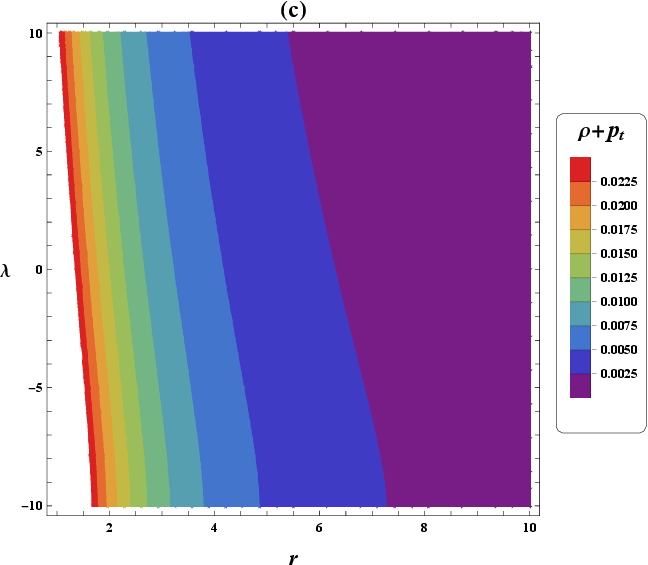}
\caption{Contour plots of the WEC for the case $C_1 = -1$. Panel $(a)$ displays the region where the energy density is valid, panel $(b)$ shows the region that satisfies $\rho + p_r$, and panel $(c)$ presents the admissible zone for $\rho + p_t$, considering different values of $\lambda$. The parameters used are: $a_1 = -1$, $a_2 = 1$, $r = 5$, and $r_0 = 1$.}
\label{fig1}
\end{figure}

Based on the results from Eqs. \eqref{set_eqs_C1_-1_1}, \eqref{set_eqs_C1_-1_2}, and \eqref{set_eqs_C1_-1_3}, the regions of the parameter space where the WEC is satisfied are graphically represented. It is observed that $\rho \geq 0$ holds for $\{ \lambda \geq -21,~a_1 \leq 0,~\forall a_2 \}$, while both $\rho + p_r \geq 0$ and $\rho + p_t \geq 0$ are satisfied for $\{ \lambda \geq -21,~\forall a_1,~a_2 \geq 0 \}$. The regions in which the WEC is fulfilled are shown in the contour plots of Fig. \ref{fig1}. Furthermore, it is worth noting that similar results are obtained even when the scale factor in Eq. \eqref{a_C_-1} is considered with a negative sign.

\subsubsection*{Case III: $C_1=1$}

\begin{figure}[h!]
\centering
\includegraphics[width=0.32\textwidth]{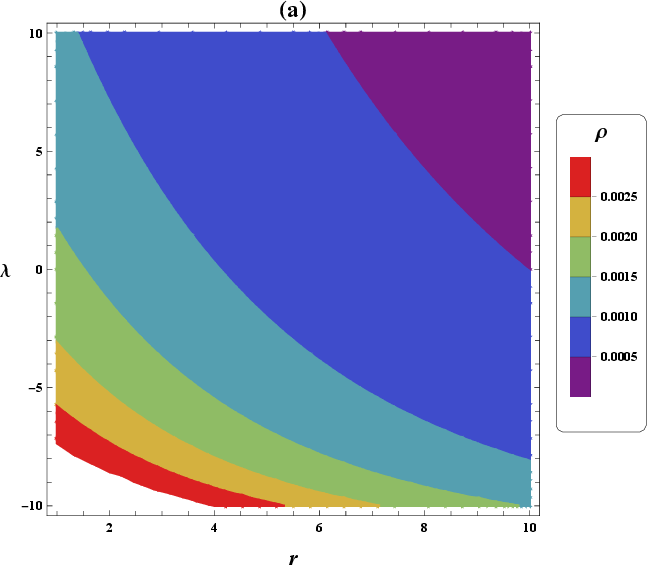}
\includegraphics[width=0.32\textwidth]{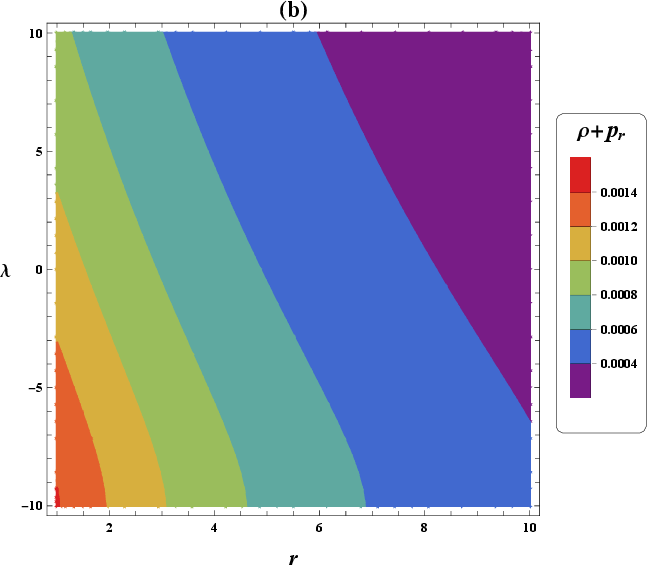}
\includegraphics[width=0.32\textwidth]{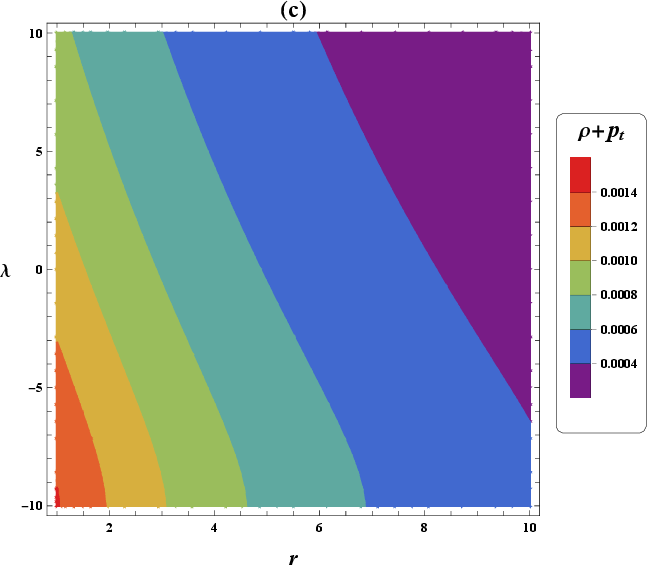}
\caption{The contour plots of the WEC for $C_1 = 1$ are shown for a specific range of $\lambda$. Plot (a) displays the region where the energy density is valid, plot (b) presents the region satisfying $\rho + p_r$ and plot (c) indicates the zone where $\rho + p_t$ holds. For this analysis, we have chosen the following values for the free parameters: $a_3 = -10$, $a_4 = 10$, $r = 5$ and $r_0 = 1$.}
\label{fig2}
\end{figure}

In the last case, we assume $C_1 = 1$ for a closed background and calculate the corresponding scale factor in order to investigate the validity of the energy constraints. By setting $\omega = -1$ in Eq. \eqref{cosmological_differential_equation}, we obtain a differential equation, the solution to which is given by:
\begin{eqnarray} \label{a_C_1}
    a(t) &=& \pm \frac{1}{\sqrt{\lambda^3 + 14 \pi \lambda^2 + 64 \pi^2 \lambda + 96 \pi^3}} \Bigg{[} (-e^{2a_3(\lambda+4 \pi)} + 144 \pi^3t^2 +84 \pi^2 \lambda t^2+16 \pi \lambda^2t^2 \nonumber \\ &+& \lambda^3 t^2+288 \pi^3 a_4 t +168 \pi^2 a_4 \lambda t+32 \pi a_4 \lambda^2t
    +a_4^2 \lambda^3+16 \pi a_4^2 \lambda^2+84 \pi^2 a_4^2 \lambda+2a_4 \lambda^3t \nonumber \\ &+& 144 \pi^3 a_4^2)\Bigg{]}^{1/2} ,
\end{eqnarray}
where $a_3$ and $a_4$ are integration constants. Using the positive expression of $a(t)$ together with $b(r)$ in Eqs. \eqref{set_ro_cb}, \eqref{set_ro_cb_1} and \eqref{set_ro_cb_12}, we identify the parameter space regions where the WEC is satisfied. Based on the graphical analysis, it is observed that $\rho \geq 0$ holds for the set $\{\lambda \geq -21,\ a_3 \leq 0,\ \forall a_4\}$, while the conditions $\rho + p_r \geq 0$ and $\rho + p_t \geq 0$ are fulfilled when $\{\lambda \geq -21,\ \forall a_3,\ a_4 \geq 0\}$. The valid regions for the WEC are illustrated in the contour plots of Fig. \ref{fig2}. It is also worth noting that similar results are obtained even when the scale factor is considered with a negative sign.

\section{Wormhole solution using anisotropic Equation of State} \label{WHs_ES}

In this section, our goal is to investigate the geometry of a static wormhole. To this end, we set $a(t) = 1$ in Eq. \eqref{form_WH_metric} and, from Eq. \eqref{final_field_equations}, we obtain:
\begin{eqnarray}\label{trace_field_eq_static}
    -R &=& (8 \pi+\lambda)T+2\lambda(\rho-P).
\end{eqnarray}
Now, by adopting an anisotropic equation of state of the form $p_t(r)=\omega p_r(r)$ \cite{Moraes:2017dbs}, we obtain the following differential equation:
\begin{eqnarray}\label{diffe_eq_static_1}
    \frac{2b'(r)}{r^2}&=&(8\pi+3\lambda)\rho-\frac{(1+2\omega)(24\pi+5\lambda)}{3}p_r.
\end{eqnarray}
Alternatively, by considering Eqs. \eqref{set_ro_pre} and \eqref{set_ro_pre_1}, we can rewrite Eq. \eqref{diffe_eq_static_1} in the following form: $b(r)$ which
is given by
\begin{eqnarray}\label{diffe_eq_static_2}
    (5 \lambda+24 \pi)(b'r(\lambda+2 \lambda \omega+12 \pi)-3b(\lambda+4 \pi)(2 \omega+1)) = 0.
\end{eqnarray}

\begin{figure}[h!]
    \centering
    \includegraphics[width=0.35\textwidth]{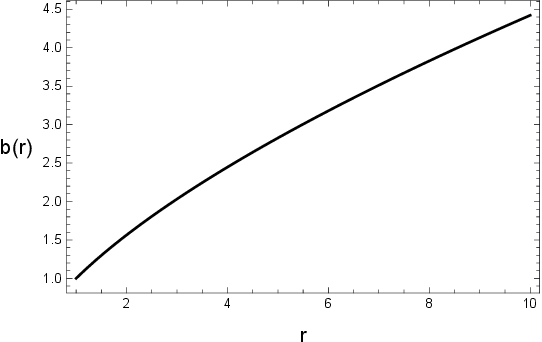}
    \includegraphics[width=0.35\textwidth]{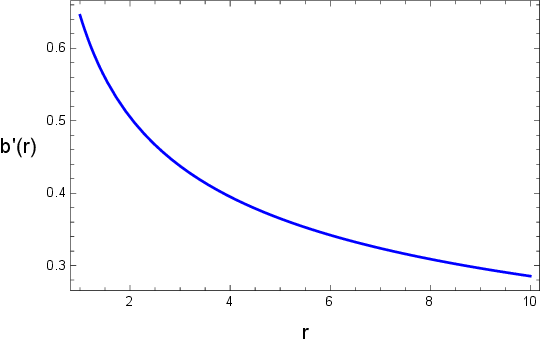}
    \includegraphics[width=0.35\textwidth]{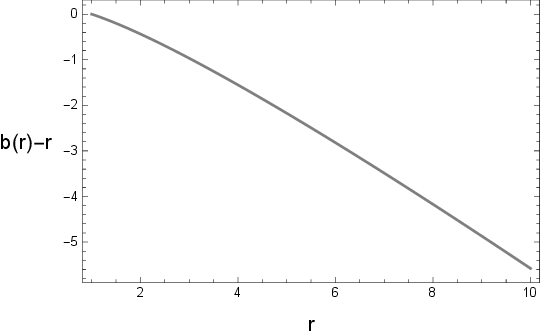}
    \includegraphics[width=0.35\textwidth]{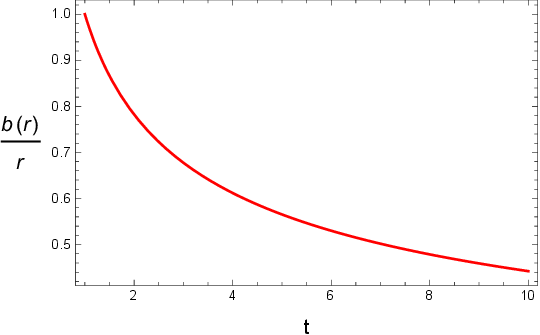}
    \caption{Behavior of the main properties of the wormhole shape function as a function of the radial coordinate $r$. The panels display: (a) the shape function $b(r)$, (b) its derivative $b'(r)$, (c) the difference $b(r) - r$, and (d) the ratio $b(r)/r$. The parameters used are $\lambda = 15$, $\omega = -1/3$, and $r_0 = 1$.}
    \label{fig6}
\end{figure}

\begin{figure}[h!]
    \centering
    \includegraphics[width=0.35\textwidth]{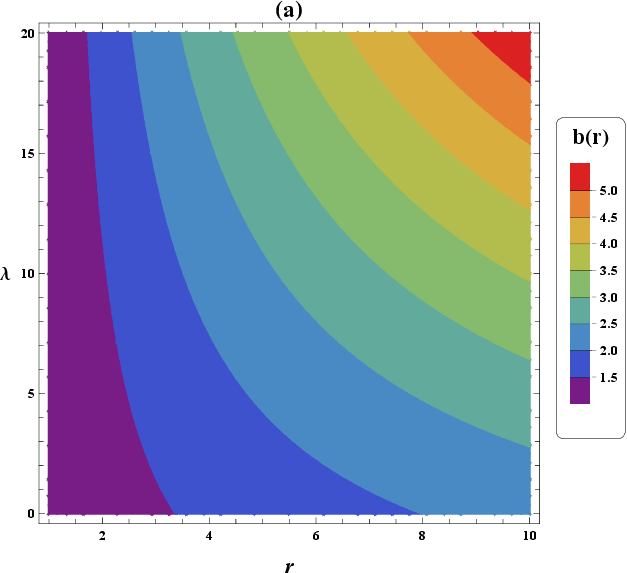}
    \includegraphics[width=0.35\textwidth]{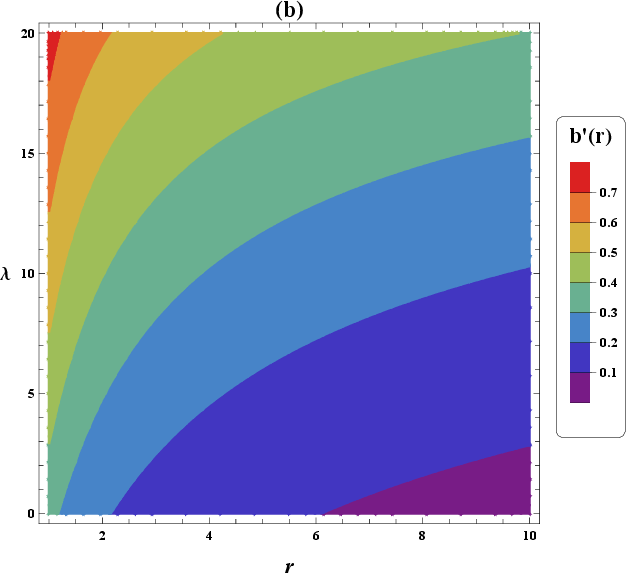} \\
    \includegraphics[width=0.35\textwidth]{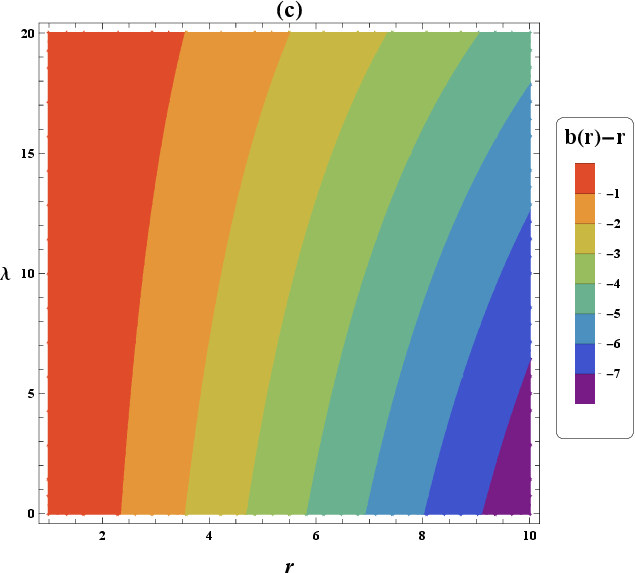}
    \includegraphics[width=0.35\textwidth]{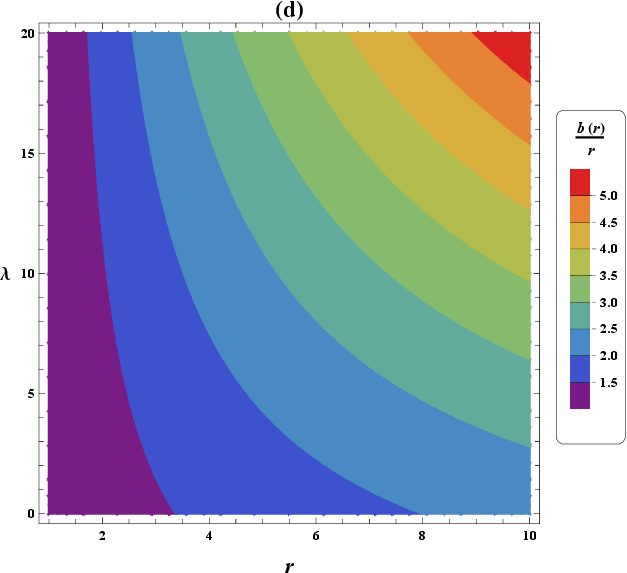}
    \caption{Behavior of the dynamics of the shape function $b(r)$ varying the parameters $r$ and $\lambda$.}
    \label{figc}
\end{figure}

\begin{figure}[h!]
    \centering
    \includegraphics[width=0.4\textwidth]{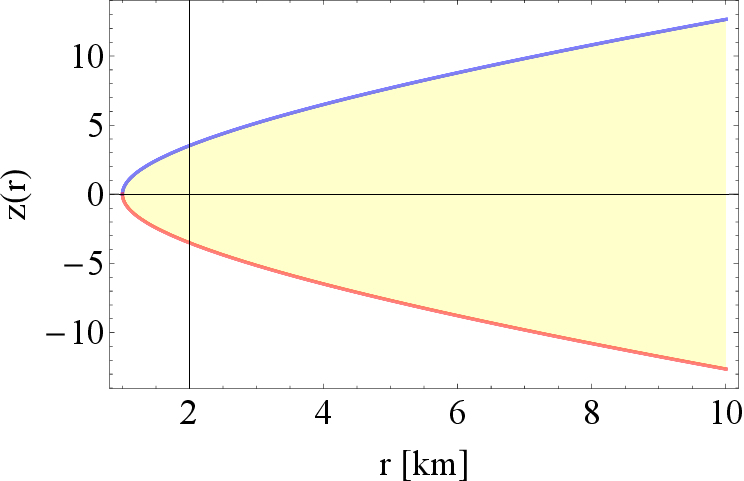}
    \includegraphics[width=0.4\textwidth]{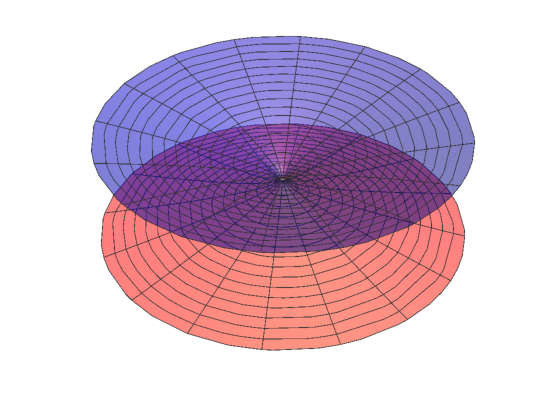}
    \caption{Embedding diagrams: the left plot shows the behavior of the shape function, while the right plot provides a complete view of the embedded curve around the $z$-axis.}
    \label{fige2}
\end{figure}

We obtain the expression for $b(r)$ by analytically solving Eq. \eqref{diffe_eq_static_2}, the solution of which is given by
\begin{eqnarray}\label{b_static}
    b(r) = b_1 r^{\frac{3(\lambda+4 \pi)(2 \omega+1)}{\lambda+2 \lambda \omega+12 \pi}},
\end{eqnarray}
where $b_1$ is an integration constant. By applying the condition $b(r_0) = r_0$, we can determine the value of $b_1$ as: 
\begin{eqnarray}
    b_1 = r_0^{1-\frac{3(\lambda+4\pi)(2\omega+1)}{\lambda+2\lambda\omega+12\pi}},
\end{eqnarray}
and therefore, $b(r)$ is given by:
\begin{eqnarray}\label{b_static_1}
    b(r) &=& r_0 \Big(\frac{r}{r_0}\Big)^{\frac{3(\lambda+4 \pi)(2 \omega+1)}{\lambda+2\lambda \omega+12 \pi}}.
\end{eqnarray}
From Eq. \eqref{b_static_1}, we can now analyze the behavior of the wormhole shape function in order to verify whether it satisfies the necessary conditions for a physically viable solution. The plots in Fig. \ref{fig6} show that the obtained function $b(r)$ exhibits behavior consistent with the criteria required for wormhole formation. The Fig. \ref{figc} displays contour plots illustrating the properties of the function $b(r)$ for different values of the parameters $\lambda$ and $r$. Finally, Fig. \ref{fige2} provides a complete visualization of the static wormhole geometry, generated by rotating the embedded curve.

\begin{figure}[h!]
\centering
\includegraphics[width=0.32\textwidth]{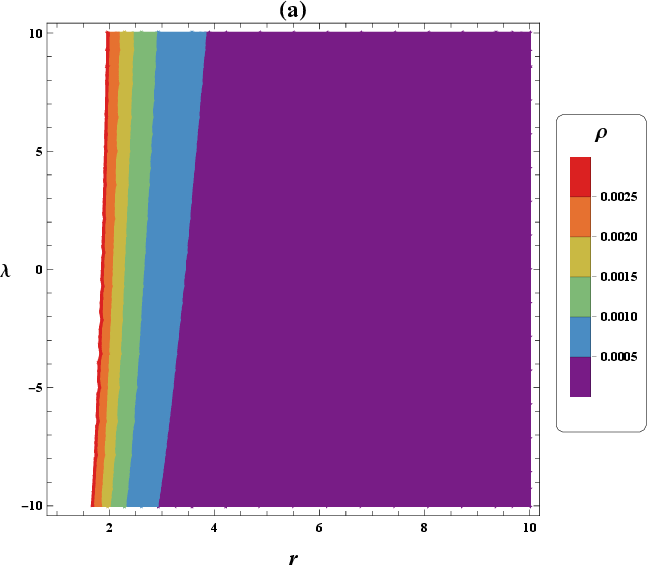}
\includegraphics[width=0.32\textwidth]{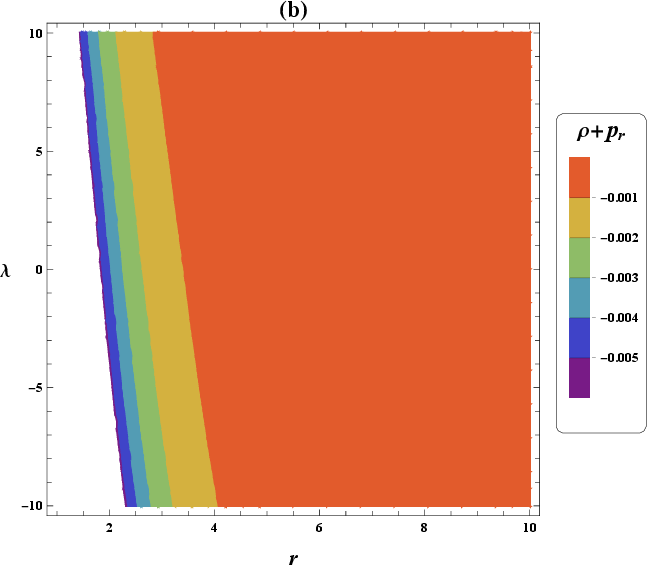}
\includegraphics[width=0.32\textwidth]{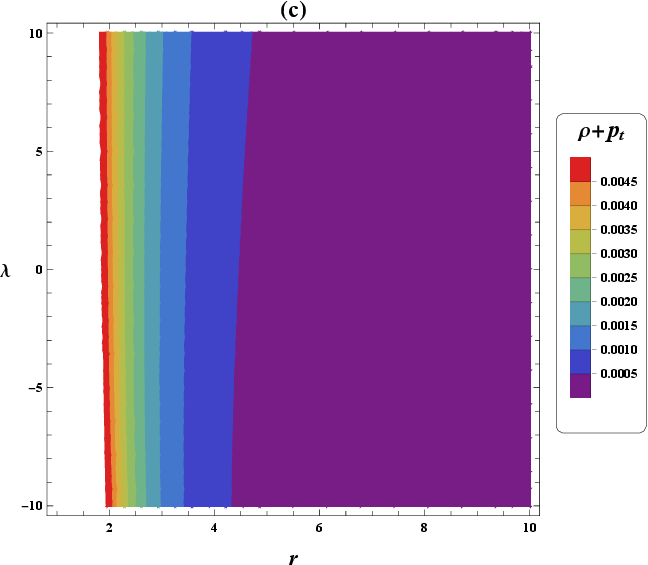}
\caption{Behavior of the WEC for the static anisotropic wormhole. The left plot shows the behavior of $\rho$, the middle plot shows $\rho + p_r$, and the right plot indicates the behavior of $\rho + p_t$. The parameters used are $\omega = -1/3$ and $r_0 = 1$.}
\label{fig7}
\end{figure}

Finally, we present the graphs in Fig. \ref{fig7} for the NEC in the wormhole geometry as characterized by Eq. \eqref{b_static_1}. We also evaluate the parameter ranges for the valid zone of the WEC and verify that all three inequalities, i.e., $\rho \geq 0$, $\rho + p_r \geq 0$ and $\rho + p_t \geq 0$, remain valid for $\left \{\lambda \geq 2, \forall \omega\right \}$.

\section{Wormhole solution using a specific energy density profile} \label{WH_EDP}

In this section, we adopt a specific form of the energy density in order to determine the shape function $b(r)$ associated with a static, spherically symmetric metric, i.e., with $a(t)=1$. This energy density profile is defined in \cite{Lobo:2012qq} and is given by:
\begin{eqnarray} \label{energy_dens_wh_static}
    \rho(r) = \rho_0(\frac{r_0}{r})^{\beta},
\end{eqnarray}
where $\rho_0$ and $\beta$ are constants. By substituting the energy density $\rho(r)$ given by Eq. \eqref{set_ro_pre}, we obtain:
\begin{eqnarray} \label{b_dens_prof}
    b(r) &=& b_0 + \frac{3(\lambda^2+12\pi \lambda+32 \pi^2) \rho_0r^3(\frac{r_0}{r})^{\beta}}{2(3-\beta)(\lambda+6 \pi)} .
\end{eqnarray}
In this case, $b_0$ is a constant of integration, which can be determined from the condition $b(r_0)=r_0$. With this consideration, we obtain:
\begin{eqnarray} \label{b0_constant}
    b_0 &=& r_0-\frac{3r_0^3 (32\ pi^2+12 \pi \lambda+\lambda^2) \rho_0}{2(3-\beta)(6 \pi+\lambda)}.
\end{eqnarray}

\begin{figure}[h!]
    \centering
    \includegraphics[width=0.35\textwidth]{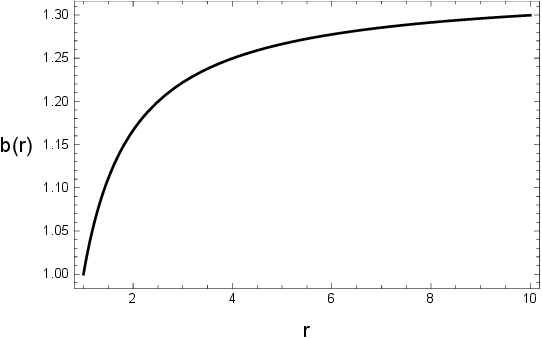}
    \includegraphics[width=0.35\textwidth]{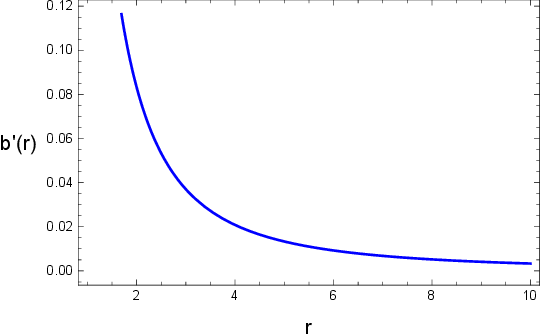}
    \includegraphics[width=0.35\textwidth]{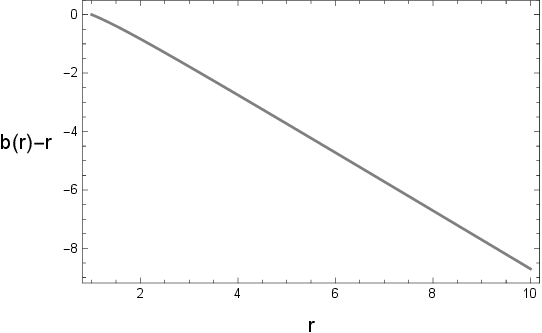}
    \includegraphics[width=0.35\textwidth]{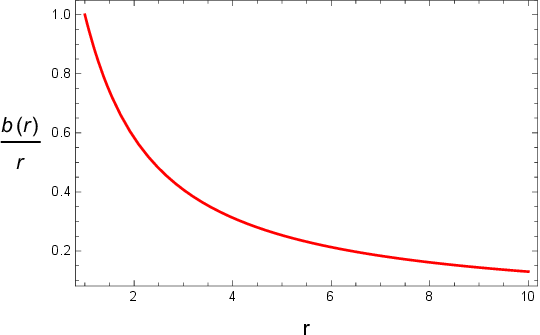}
    \caption{Behavior of the shape function $b(r)$ characteristics. At the top, the first plot shows the variation of $b(r)$, while the second displays its derivative $b'(r)$. At the bottom, the first plot presents the difference $b(r)-r$, and the second shows the ratio $b(r)/r$. For this analysis, we adopt the following parameter values: $\lambda=5$, $\rho=0.01$, $\beta=4$ and $r_0=1$.}
    \label{fig8}
\end{figure}

\begin{figure}[h!]
    \centering
    \includegraphics[width=0.35\textwidth]{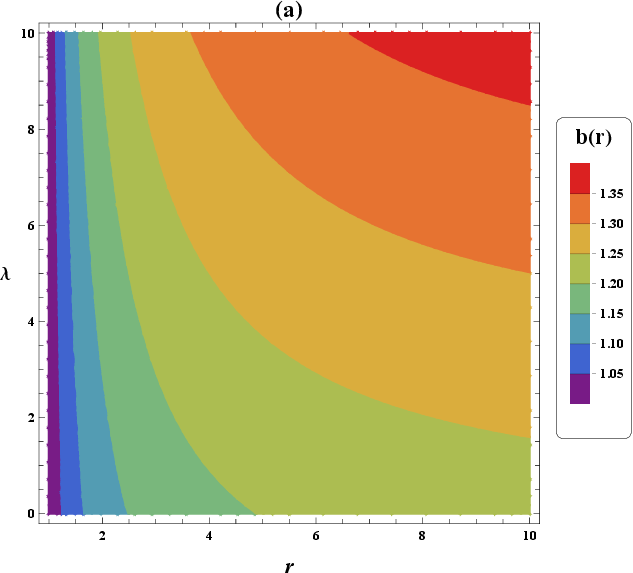}
    \includegraphics[width=0.35\textwidth]{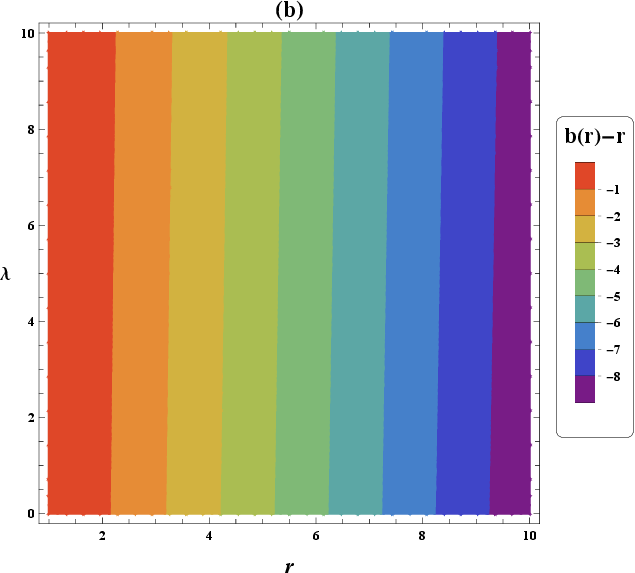}
    \caption{Dynamical behavior of the shape function as a function of $r$, considering a specific range of $\lambda$. The remaining parameters are fixed as $\rho_0 = 0.01$, $\beta = 4$ and $r_0 = 1$.}
    \label{fig8*}
\end{figure}

\begin{figure}[h!]
    \centering
    \includegraphics[width=0.4\textwidth]{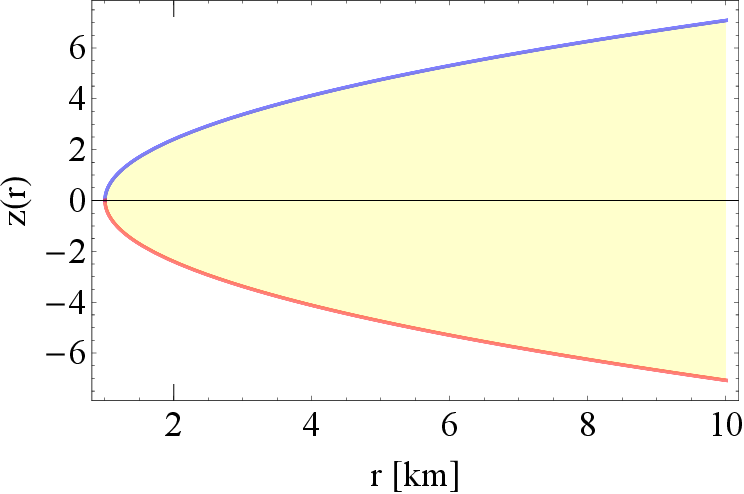}
    \includegraphics[width=0.4\textwidth]{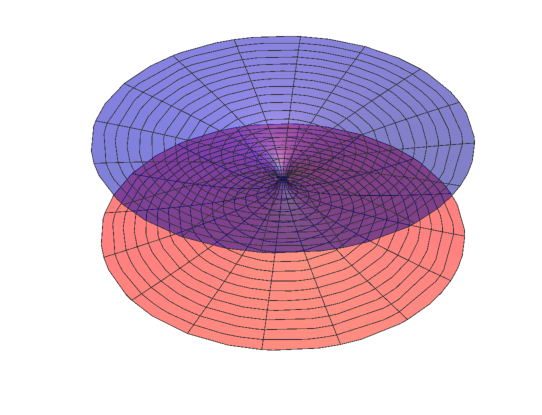}
    \caption{The behavior of the embedding diagram of the wormhole is shown in the first plot. The second plot provides a complete view of the embedded curve around the $z$-axis. The parameters used are $\lambda = 5$, $\rho_0 = 0.01$, $\beta = 4$, and $r_0 = 1$, respectively.}
    \label{fig8**}
\end{figure}

Therefore, by substituting Eq.\eqref{b0_constant} into Eq.\eqref{b_dens_prof}, the final form of the WH shape function is determined. The behavior of $b(r)$ is shown in Fig. \ref{fig8}, along with the verification of the fundamental axioms that this function must satisfy. It is observed that the proposed wormhole model complies with all these criteria, as also illustrated in Fig. \ref{fig8*}. Furthermore, the embedding diagram associated with the shape function $b(r)$ is displayed in Fig. \ref{fig8**}.

\begin{figure}[h!]
    \centering
    \includegraphics[width=0.32\textwidth]{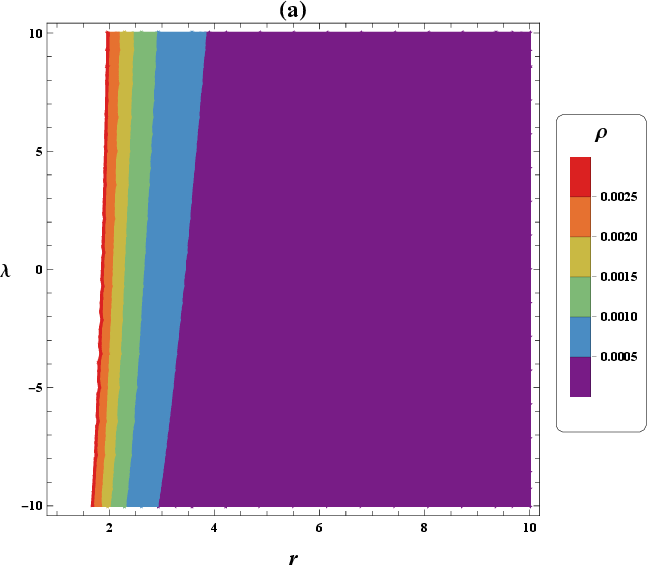}
    \includegraphics[width=0.32\textwidth]{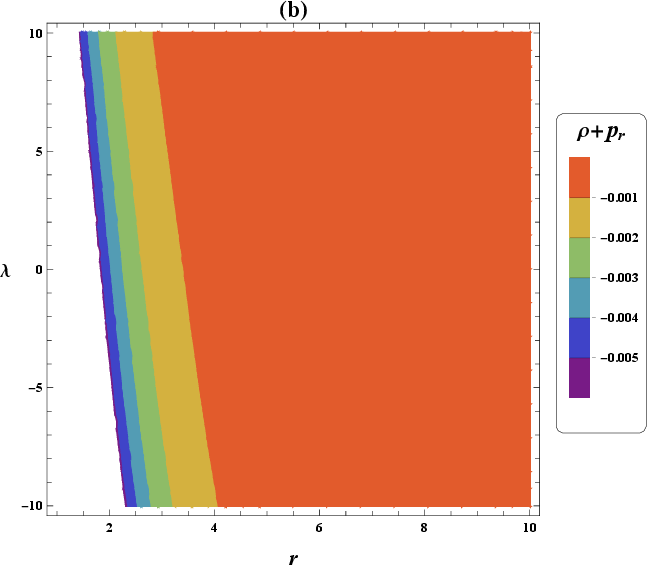}
    \includegraphics[width=0.32\textwidth]{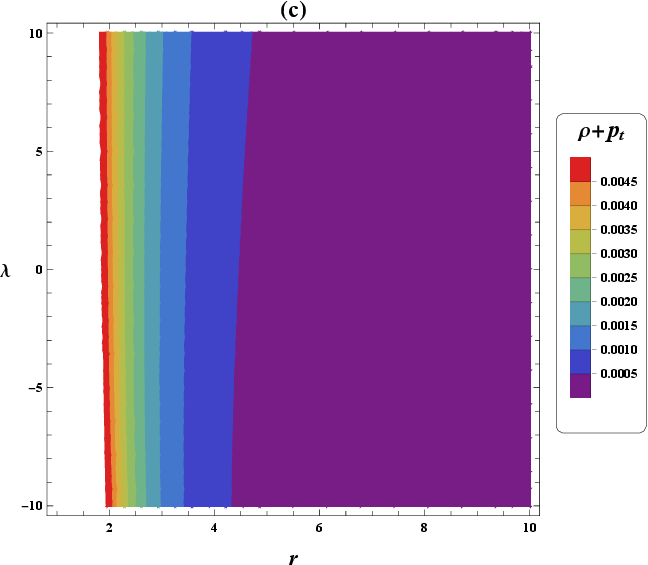}
    \caption{Behavior of the WEC for the wormhole given by Eq. \eqref{b_dens_prof}. The left plot shows the trend of $\rho$, the middle plot illustrates the behavior of $\rho + p_r$, while the right plot shows the behavior of $\rho + p_t$. The parameters are fixed as $\rho_0 = 0.01$, $\beta = 4$, and $r_0 = 1$.}
    \label{fig9}
\end{figure}

Finally, we graphically identify the zones where the WEC can be satisfied. It is found that $\rho \geq 0$ holds for $\left \{ \forall \lambda, \forall \beta \right \}$. Moreover, $\rho + p_r \geq 0$ is valid when $\left \{\lambda \leq -26, \forall \beta\right \}$, and $\rho + p_t \geq 0$ remains valid for $\left \{\lambda \geq 0, \forall \beta\right \}$. The contour plots illustrating the validity of the WEC for specific values of $\lambda$ are presented in Fig. \ref{fig9}.

\section{Equilibrium Condition} \label{equi_cond}

To analyze the stability of the solutions obtained in the previous sections, this section investigates the equilibrium condition by means of the generalized Tolman-Oppenheimer-Volkoff (TOV) equation, applied to each wormhole configuration considered. Mathematically, the generalized TOV equation is expressed as \cite{Kuhfittig:2013hva}:
\begin{eqnarray} \label{t}
    -\frac{dp_r}{dr}-\frac{\Phi'(r)}{2}(\rho+p_r)-\frac{2}{r}(p_r-p_t)=0 .
\end{eqnarray}
In the equation above, the different terms represent distinct forces acting on the WH configuration and can be identified as follows:
\begin{eqnarray}
    F_h=-\frac{dp_r}{dr}, \ \ \ \ F_g=-\frac{\Phi'(r)}{2}(\rho+p_r), \ \ \ \ F_a=\frac{2(p_t-p_r)}{r} ,
\end{eqnarray}
where $F_h$, $F_g$, and $F_a$ represent the hydrostatic, gravitational, and anisotropic forces, respectively. Thus, the equilibrium condition can be expressed as follows:
\begin{eqnarray}
    F_h+F_g+F_a=0.
\end{eqnarray}

\begin{figure}[h!]
    \centering
    \includegraphics[width=0.4\textwidth]{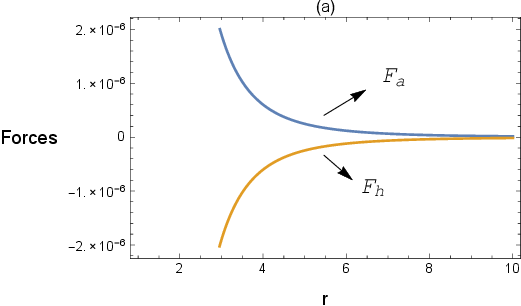}
    \includegraphics[width=0.4\textwidth]{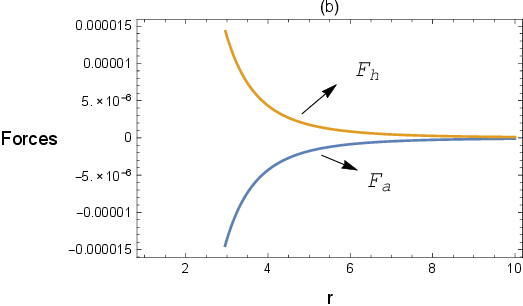}
    \caption{Dynamic behavior of hydrostatic and anisotropic forces for $b(r)=C_1r^3-r_0(C_1r_0^2-1)$, with $C_1=0$ and $C=-1$ in the left and right panels, respectively. The fixed parameters are $A=10$, $B=10$, $a_1=-10$, $a_2=10$, $\lambda=10$, and $r_0=1$.}
    \label{fig10}
\end{figure}

\begin{figure}[h!]
    \centering
    \includegraphics[width=0.4\textwidth]{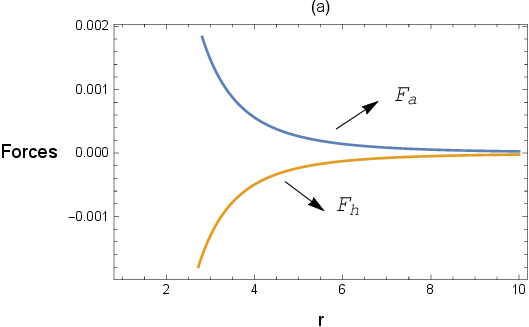}
    \includegraphics[width=0.4\textwidth]{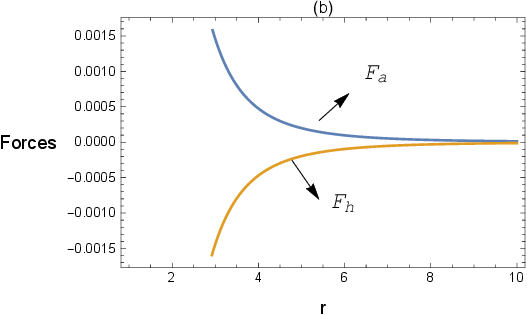}
    \caption{The dynamics of the forces from Eqs. \eqref{b_static_1} and \eqref{b_dens_prof} are illustrated below. The left plot shows the equilibrium configuration for the shape function $b(r)$ from Eq. \eqref{b_static_1}, while the right plot presents the equilibrium configuration for the shape function from Eq. \eqref{b_dens_prof}. The chosen parameters are: $\rho_0 = 0.01$, $\beta = 4$, $\lambda = 15$, $\omega = -1/3$, and $r_0 = 1$.}
    \label{fig11}
\end{figure}

Since the redshift function is constant, i.e., $\Phi'(r) = 0$, the gravitational force becomes null, and therefore the equilibrium condition simplifies to:
\begin{eqnarray}
    F_h + F_a = 0.
\end{eqnarray}
Thus, we can calculate the hydrostatic and anisotropic forces using the $b(r)$ functions of the wormholes obtained from the Ricci scalar, the anisotropic equation of state, and the energy density profile, as given by Eqs. \eqref{scale_R_r_t}, \eqref{b_static_1}, and \eqref{b_dens_prof}, respectively. Thus, we can plot the behavior of these forces, as shown in Figs. \ref{fig10} and \ref{fig11}, for each case of $b(r)$. The left plot in Fig. \ref{fig10} illustrates the behavior of the forces for the shape function specified in Eq. \eqref{scale_R_r_t} with $C_1 = 0$, while the right plot shows the behavior for $C_1 = -1$. Similarly, Fig. \ref{fig11} presents the behavior of the forces for the shape functions derived from the anisotropic EOS and the energy density profile, respectively. It is clear that each of these forces exhibits the same behavior, but in opposite directions, resulting in the cancellation of their net effects, thus indicating stable wormhole geometries.

\section{Complexity factor for dynamical wormholes in $f(R,T)$ gravity} \label{y_factor}

The complexity factor, $Y_{TF}$, provides a valuable tool for understanding how fluctuations in energy density and pressure influence the structure and dynamics of gravitational systems. This concept was initially introduced within the framework of GR \cite{Herrera:2018bww}, offering a way to quantify the complexity of self-gravitating systems, particularly those that are spherically symmetric and static. Later, the idea was extended to include dynamical self-gravitating systems \cite{Herrera:2018czt}.

More recently, the $Y_{TF}$ has been investigated in the context of modified gravity, specifically for spherically symmetric, dissipative, dynamical self-gravitating systems in $f(R,T)$ gravity \cite{Zubair:2020nmx}. In this scenario, by considering Eqs. \eqref{field_equations} and \eqref{form_WH_metric}, the complexity factor takes the following form:
\begin{eqnarray}\label{ytf1}
    Y_{TF} = \frac{3}{2 a^3 r^3}\int a^3 r^2 \left(\rho-\frac{1}{2}\lambda T\right)dr+\frac{1}{4}\lambda T+(\lambda+1)\Pi-\frac{\rho }{2},
\end{eqnarray}
where $\Pi = p_r - p_t$ represents the pressure anisotropy, and $T$ is the trace of the energy-momentum tensor. Based on Eqs. \eqref{set_ro_pre}, \eqref{set_ro_pre_1} and \eqref{set_ro_pre_2}, the complexity factor is given by:
\begin{eqnarray} \nonumber
    Y_{TF} &=& \frac{1}{12a^2(\lambda+4 \pi)(\lambda+8 \pi)r^3} \Bigg\{\lambda \Psi r \left(3a^2H'r^2+2b'(r)\right)+9a^2 H^2(\lambda+8 \pi)(2 \lambda-1)r^3 \\ &-& 18(\lambda+1)(\lambda+4 \pi) b(r) + \frac{3}{a} \int_{r_0}^{r} \Big[3a^3 r^2 \left(-3H^2(\lambda+8 \pi)(2 \lambda-1) -H' \lambda \Psi \right) \label{ytf2} \\ &-& 2a (12 \pi(\lambda-1)
    +(\lambda-2)\lambda)b'(r)\Big]dr\Bigg\}, \nonumber
\end{eqnarray}
where we define $\Psi = (4\lambda + 24\pi + 1)$. We now proceed to calculate the complexity factors by incorporating the redshift function and the scale factor for the different WHs cases discussed in the previous sections.

\subsubsection*{Cosmological Wormholes: $C_1=0$}

\begin{figure}[h!]
    \centering
    \includegraphics[width=0.4\textwidth]{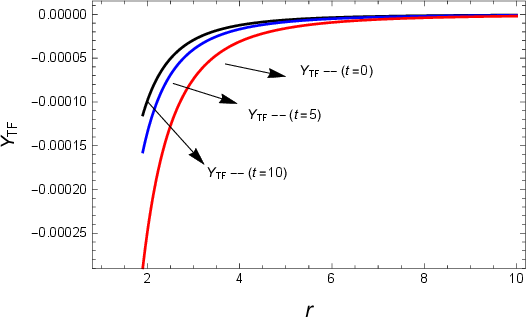}
    \includegraphics[width=0.4\textwidth]{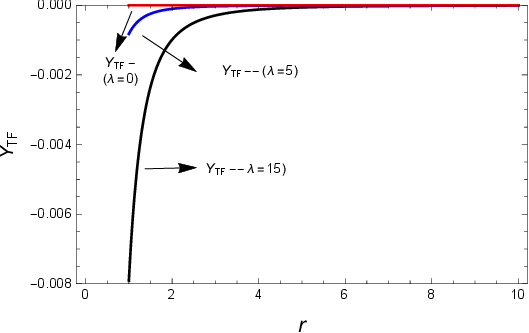}
    \caption{Evolution of the complexity factor $Y_{TF}$ for a cosmological wormhole with $C_1 = 0$. The left panel shows an initially negative complexity, whose magnitude decreases over time. The right panel displays the evolution of $Y_{TF}$ for different values of the parameter $\lambda$, highlighting the influence of this parameter on the system's structural complexity.}
    \label{fy1}
\end{figure}

We begin by considering the parameters adopted in the case of a cosmological wormhole. By substituting Eqs. \eqref{scale_R_r_t} and \eqref{cosmological_differential_equation}, we obtain the following expression for the complexity factor:
\begin{eqnarray} \label{y1}
    Y_{TF} &=& \frac{\lambda(4 \lambda+24 \pi+1)
    \left(3C_1r^3-3 \left(C_1 r^3-r_0 \left(C_1 r_0^2-1 \right) \right) \right)}{6B^2(\lambda+4 \pi)(\lambda+8 \pi)r^3(A+3 t)^{2/3}}.
\end{eqnarray}
From the plots in Fig. \ref{fy1}, we observe the variations of the complexity factor $Y_{TF}$ as a function of the radial coordinate $r$, for different time values and different values of the parameter $\lambda$ in the modified gravity theory $f(R, T)$. In the left panel, the system’s initial complexity is significantly negative, and its magnitude decreases over time. The negative sign of $Y_{TF}$ in the context of wormholes indicates that the contributions from anisotropic pressure outweigh the effects of energy density inhomogeneity, effectively simplifying the global structure of the system.

Although the complexity factor approaches zero in the distant future, it remains negative and does not become positive. Furthermore, at large radial distances, the complexity factor vanishes at all times, reaching its maximum magnitude near the wormhole throat. This behavior suggests that under certain conditions, deviations from uniformity and isotropy can reduce the perceived complexity of the system.

In the right panel, the evolution of $Y_{TF}$ is shown for different values of $\lambda$, revealing a remarkable behavior: when $\lambda = 0$, the complexity factor remains zero at all times. As $\lambda$ increases, the magnitude of $Y_{TF}$ also increases, indicating a direct relationship between this parameter and the structural complexity of the system.

\subsubsection*{Cosmological Wormholes: $C_1= -1, 1$}

Next, we analyze the complexity factor in two distinct scenarios:
\begin{itemize}
    \item For a wormhole in an open background, $C_1 = -1$, with a scale factor from an inflationary expansion regime, $\omega = -1$,
    \item For a wormhole in a closed background ($C_1 = 1$), with a scale factor corresponding to the inflationary expansion regime ($\omega = -1$).
\end{itemize}

\begin{figure}[h!]
    \centering
    \includegraphics[width=0.38\textwidth]{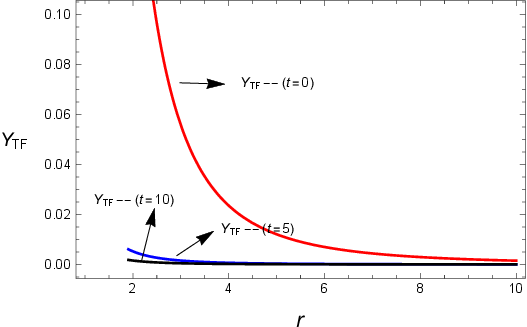}
    \includegraphics[width=0.38\textwidth]{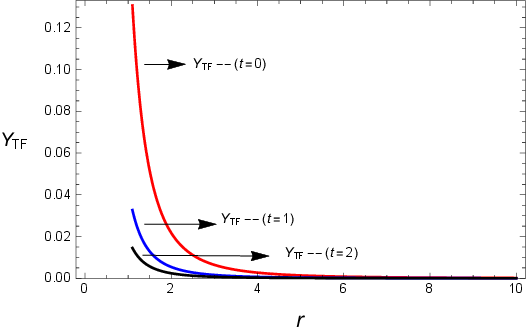}
    \includegraphics[width=0.38\textwidth]{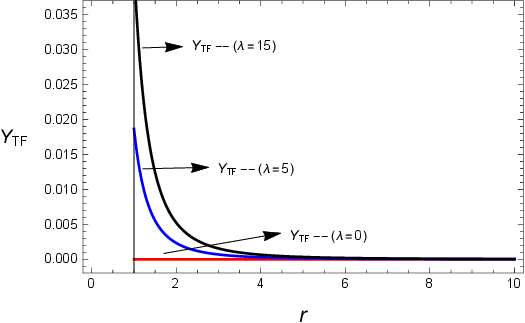}
    \caption{Behavior of the complexity factor for wormholes with $C_1=-1$ and $C_1=1$. In both scenarios, the factor $Y_{TF}$ exhibits the same form, indicating an identical evolution of the structural complexity.}
    \label{fy2}
\end{figure}

Since both scenarios result in identical complexity factors, we consider them together in this subsection. Thus, by substituting Eqs. \eqref{scale_R_r_t} and \eqref{a_C_-1} into Eq. \eqref{ytf2}, we obtain the following expression for $Y_{TF}$:
\begin{eqnarray} \label{y2}
    Y_{TF} &=& -\frac{\lambda(\lambda+4 \pi)(\lambda+6 \pi)(4 \lambda+24 \pi+1)r_0 \left(C_1r_0^2-1\right)}{2(\lambda+8 \pi)r^3 \left((\lambda+4 \pi)(\lambda+6 \pi)^2(a_2+t)^2-e^{2a_1(\lambda+4 \pi)}\right)}.
\end{eqnarray}
To better understand the evolution of complexity in the two WHs scenarios, we present the plots in Fig. \ref{fy2}. The fact that the complexity factor $Y_{TF}$ is positive indicates that the internal structure of the WHs exhibits greater complexity and reduced symmetry. This suggests that such complexity arises from factors like non-uniform energy density and anisotropic pressure, whose effects do not fully counterbalance each other.

The left and right plots in the figure illustrate the temporal influence on this complexity factor. Initially, the WHs display significant deviations from uniform energy density and pressure isotropy. As a result, this higher degree of complexity may affect the stability of the WH, potentially making it more sensitive to fluctuations or requiring more restrictive conditions to maintain its structure.

However, as time progresses, the complexity factor tends to decrease, indicating that the system evolves toward a more stable or simplified configuration. The bottom plot in Fig. \ref{fy2} shows the variations in the complexity factor for different values of $\lambda$. For instance, when $\lambda = 0$, the complexity factor $Y_{TF}$ is zero; as $\lambda$ increases, the complexity also rises, revealing greater sensitivity to perturbations. In all scenarios, the complexity factor reaches its highest values near the throat of the WHs and tends toward zero as one moves away from it.

\subsubsection*{Wormholes with Anisotropic EOS and Energy Density Profile}

Finally, we analyze the variations of the complexity factor for wormhole solutions derived from an anisotropic equation of state and a specific energy density profile, as previously discussed. To do so, we substitute the redshift function given in Eq. \eqref{b_static_1}, thus obtaining the following expression for $Y_{TF}$:
\begin{eqnarray} \label{y3}
    Y_{TF} &=& \frac{\lambda(4 \lambda+24 \pi+1)\left(\frac{3(\lambda+4 \pi) \Psi_2 r(2w+1)}{\lambda+2 \lambda w+12 \pi}-3\Psi_1 r_0\right)}{6(\lambda+4 \pi)(\lambda+8 \pi)r^3},
\end{eqnarray}
where 
\begin{eqnarray}
    \Psi_1=\left(\frac{r}{r_0}\right)^{\frac{3(\lambda+4\pi)(2w+1)}{\lambda+2\lambda w+12\pi}} \ \ \ \ \text{and} \ \ \ \ \Psi_2=\left(\frac{r}{r_0}\right)^{\frac{2(\lambda+2(\lambda+6\pi)w)}{\lambda+2\lambda w+12\pi}}.
\end{eqnarray}
On the other hand, by substituting the value of the redshift given in Eq. \eqref{b_dens_prof}, we obtain the following expression for $Y_{TF}$:
\begin{eqnarray} \label{y4}
    Y_{TF} = \frac{\lambda(4 \lambda+24 \pi+1)}{6a^2(\lambda+4 \pi)(\lambda+8 \pi)r^3} &\Bigg\{&\frac{3(\lambda+4 \pi)(\lambda+8 \pi)\rho_0 r^3 \left(\frac{r_0}{r}\right)^{\beta}}{2(\lambda+6 \pi)} \\ &-&3\left[\frac{3\left(\lambda^2+12 \pi \lambda+32 \pi^2\right) \rho_0 \left(r_0^3-r^3 \left(\frac{r_0}{r}\right)^{\beta}\right)}{2(\beta-3)(\lambda+6 \pi)}
    +r_0 \right]\Bigg\} \nonumber.
\end{eqnarray}

\begin{figure}[h!]
    \centering
    \includegraphics[width=0.38\textwidth]{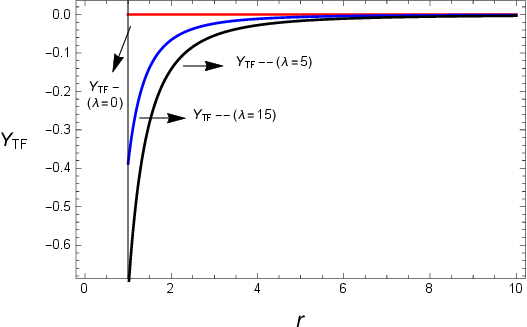}
    \includegraphics[width=0.38\textwidth]{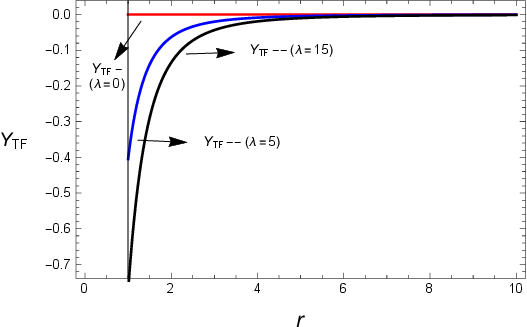}
    \caption{The figure shows the evolution of the complexity factor for wormhole solutions obtained from an anisotropic equation of state (left) and from a specific energy density profile (right).}
    \label{fy3}
\end{figure}

Thus, it is observed that in both cases, the complexity factor is zero for $\lambda = 0$, and its magnitude increases with a negative sign as $\lambda$ grows, indicating greater sensitivity to perturbations. Furthermore, the factor $Y_{TF}$ reaches higher values near the wormhole’s throat and asymptotically approaches zero at larger distances, as illustrated in Fig. \ref{fy3}.

\section{Conclusions and Perspectives} \label{conclusions}

Modern Cosmology has shown great interest in the development of viable wormhole geometries, particularly within the framework of modified gravity theories. In this context, significant efforts have been made in recent years. This interest is justified by the potential of such solutions to provide new insights into the fundamental nature of gravity, enabling the exploration of configurations that are initially exotic but may have meaningful implications for our understanding of the universe. By investigating wormholes in the realm of modified gravitational theories, researchers aim to assess their feasibility, stability, and the physical conditions required for their possible existence.

In this article, we investigate the viability of dynamic wormhole solutions supported by ordinary matter, considering three different phases of the universe within the well-formulated framework of modified gravity $f(R,T)$. This gravitational theory is particularly promising as it allows for a direct interaction between spacetime curvature and matter content. We adopt a linear model for the $f(R,T)$ function, representing the simplest possible coupling between geometry and matter.

Unlike General Relativity, in which the geometry of a wormhole is strongly constrained by the energy conditions imposed by matter, $f(R,T)$ gravity introduces an effective energy-momentum tensor that supports the geometric configuration. This makes it possible to construct solutions that are compatible with the principles of GR while offering greater flexibility to incorporate cosmological phenomena, such as the accelerated expansion of the universe and structure formation.

This work expands upon the analysis presented in \cite{Zubair:2022jjm}, extending it to the dynamical case. To this end, we consider a spherically symmetric and non-static spacetime filled with an anisotropic fluid. We determine the wormhole shape function from the Ricci scalar under a cosmological condition and analyze the scale factor for three distinct values of the integration constant $C_1$, corresponding to flat, open, and closed universe models. For each case, we examine the associated energy conditions in order to understand the evolution of dynamic wormhole configurations in $f(R,T)$ gravity.

We also assess the validity of the WEC throughout the evolution, identifying the regimes in which it is satisfied or violated. Additionally, we incorporate an anisotropic equation of state and a specific energy density profile to derive the shape function $b(r)$, aiming to construct physically viable WHs geometries. Below, we summarize the main results obtained:
\begin{itemize}
    \item Initially, we fixed the integration constant at $C_1 = 0$ (flat universe) in the shape function and evaluated the corresponding scale factor $a(t)$. We analyzed the WEC and NEC energy conditions and found that both are satisfied for $\lambda \geq -25$, considering $\omega = 1$.
    
    \item Next, we examined the case $C_1 = -1$ (open universe) with $\omega = -1$, and from the plots in Fig. \ref{fig1}, we observed that the WEC holds for $\lambda \geq -21$.

    \item Then, we considered $C_1 = 1$ (closed universe), also with $\omega = -1$, and verified that the WEC is satisfied at the wormhole throat for $\lambda \geq -21$, suggesting the possibility of dynamic wormhole solutions within this gravitational theory.

    \item Additionally, we considered an anisotropic equation of state and a specific energy density profile to obtain the shape function $b(r)$ in a static and spherically symmetric spacetime. We explored the energy conditions through graphical analysis for all considered scenarios Figs. \ref{fig6}, \ref{figc}, \ref{fige2}, \ref{fig7}, \ref{fig8}, \ref{fig8*}, \ref{fig8**} and \ref{fig9}, identifying the regions of validity.

    \item To assess the consistency and stability of the solutions, we employed the equilibrium condition from the TOV equation. As shown in Figs. \ref{fig10} and \ref{fig11}, we concluded that the acting forces are nearly equal in magnitude but opposite in direction, mutually canceling and leading to an equilibrium configuration for the wormhole geometry.

    \item Finally, we investigated the complexity factor for the WHs solutions. For $C_1 = 0$, the $Y_{TF}$ is negative near the throat, indicating that the anisotropic pressure reduces the overall complexity by outweighing the inhomogeneity in energy density. Over time, $Y_{TF}$ tends to zero but remains negative, with maximum magnitude near the throat and vanishing at larger distances. As the $\lambda$ parameter in $f(R,T)$ gravity increases, the absolute value of the complexity factor also increases, showing that $\lambda$ directly influences the system's complexity.

    \item For $C_1 = -1$ and $C_1 = 1$, $Y_{TF}$ is positive, indicating greater internal complexity and reduced symmetry due to non-uniform energy density and unbalanced anisotropic pressure. Initially, the wormhole exhibits significant deviations, but complexity decreases over time, leading to a more stable structure. As $\lambda$ increases, the positive complexity factor also increases, reflecting higher sensitivity to perturbations near the wormhole throat.

    \item For WHs solutions based on the anisotropic EoS and a specific energy density profile, $Y_{TF}$ is zero when $\lambda = 0$, but grows in negative magnitude as $\lambda$ increases, reaching a maximum near the throat and tending to zero at larger distances, indicating increased structural complexity in the system.
\end{itemize}

As a future perspective, it would be interesting to extend the study of dynamic wormholes to other modified theories of gravity. Such investigations may reveal new physically viable scenarios and provide a deeper understanding of spacetime structure in dynamic contexts, thus contributing significantly to the development of alternative theories to General Relativity.



\end{document}